\documentclass[%
reprint,
aps,
prb,
10pt,
twocolumn,
amsmath,
amssymb]{revtex4-2}

\usepackage{graphicx}
\usepackage{dcolumn}
\usepackage{bm}
\usepackage[utf8]{inputenc}
\usepackage[T1]{fontenc}
\usepackage{mathptmx}
\usepackage{etoolbox}
\usepackage{booktabs}
\usepackage{array} 
\usepackage[version=4]{mhchem}

\usepackage[dvipsnames]{xcolor}

\usepackage{siunitx, tabularx}    
\begin{document}

\title{Investigation of O interstitial diffusion in $\beta$-Ga$_2$O$_3$: direct approach via master diffusion equations}

\author{Grace McKnight}
    \affiliation
    {Department of Mechanical Science and Engineering, University of Illinois at Urbana-Champaign, 1206 W. Green Street, Urbana, Illinois 61801, United States.}
\author{Channyung Lee}
    \affiliation
    {Department of Mechanical Science and Engineering, University of Illinois at Urbana-Champaign, 1206 W. Green Street, Urbana, Illinois 61801, United States.}
\author{Elif Ertekin}
    \affiliation
    {Department of Mechanical Science and Engineering, University of Illinois at Urbana-Champaign, 1206 W. Green Street, Urbana, Illinois 61801, United States.}
    \affiliation
    {Materials Research Laboratory, University of Illinois at Urbana-Champaign, Urbana, Illinois 61801}
    \email{ertekin@illinois.edu}

\begin{abstract}
Monoclinic $\beta$-Ga$_2$O$_3$, a promising wide band gap semiconducting material, exhibits complex, anisotropic diffusional characteristics and mass transport behavior as a results of its low symmetry crystal structure.
From first-principles calculations combined with master diffusion equations, we determine three-dimensional diffusion tensors for neutral ($\text{O}_{\text{i}}^{0}$) and 2- charged oxygen interstitials ($\text{O}_{\text{i}}^{2-}$).
Systematic exploration of the configurational space identifies stable configurations in these two dominant charge states and their corresponding formation energies.
By connecting every pair of low-energy configurations considering both interstitial or interstitialcy hops, we construct three-dimensional diffusion networks and evaluate hopping barriers of all transition pathways in networks.
Combining the collection of (i) defect configurations and their formation energies and (ii) the hopping barriers that link them, we construct and solve the master diffusion equations for $\text{O}_{\text{i}}^{0}$ and $\text{O}_{\text{i}}^{2-}$ separately through the Onsager approach, resulting in respective three-dimensional diffusion tensors D$_{\text{O}_{\text{i}}}^{0}$ and D$_{\text{O}_{\text{i}}}^{2-}$.
Both $\text{O}_{\text{i}}^{0}$ and $\text{O}_{\text{i}}^{2-}$ present the fastest diffusion along the $b$-axis, demonstrating significant anisotropy.
The predicted self-diffusivities along [100] and [$\overline{2}01$] align well with previously reported values from isotopically labeled oxygen tracer experiments, highlighting the reliability of the approach in capturing complex diffusion mechanisms.
\end{abstract}

\maketitle

\section{\label{sec:level1}Introduction}
Beta gallium oxide ($\beta$-Ga$_2$O$_3$) has garnered interest as a wide band gap semiconductor due to its large band gap of 4.8 eV \cite{Tippins1965, Matsumoto_1974, Ueda1997}, high breakdown electric field \cite{Shigenobu2012}, and notable thermal stability \cite{Kazuo2008, Bickermann2014}. 
These unique properties make it ideal for high-power and high-frequency electronic devices, such as power transformers, UV photo-detectors, solar cells, and sensors subject to extreme conditions \cite{GreenAIP2022, Pearton2018, HuangACS2024, Perez2019, Giri2024}. 
Advances in the synthesis of high-quality $\beta$-Ga$_2$O$_3$ single crystal wafers \cite{Bickermann2014, Aida_2008, Weihua2024} and thin-films \cite{Takeuchi2008, Higashiwaki2014, Orita2000} have further enhanced its potential. 
Compared to its predecessors, silicon carbide (SiC) and gallium nitride (GaN), $\beta$-Ga$_2$O$_3$ offers distinct advantages, positioning it as a crucial material for future power electronics and optoelectronic applications.

In semiconducting materials of interest for power electronics like $\beta$-Ga$_2$O$_3$, realizing their full potential requires a detailed understanding of intrinsic defect diffusion, as it plays a crucial role in determining electrical performance and device stability.
In particular, the low-symmetry monoclinic crystal structure of $\beta$-Ga$_2$O$_3$ presents multiple diffusion pathways and directionally distinct defect interactions, making defect transport mechanisms especially rich and challenging to understand. 
Failing to control defect migration can cause unintended mass transfer, phase segregation, defect clustering, and reduced reliability \cite{Motti2021, Kabir2017, Jeong2013}.
A thorough investigation of native diffusion behavior lays the groundwork for precisely tailoring material properties to suit device-specific applications. 
This is especially important in anisotropic semiconductors, where native defects significantly influence direction-dependent properties.

Oxygen ion diffusion in semiconducting oxides typically occurs via either vacancy- or interstitial-mediated mechanisms.
Although most oxides demonstrate vacancy-dominated O transport, materials characterized by O$_\text{i}$ transport exhibit exceptionally low diffusion activation energies (E$_\text{act}$) \cite{Meng2024}.
Computational studies of $\beta$-Ga$_2$O$_3$ predicted barriers as low as 0.14 eV for O$_\text{i}$ \cite{ 18_Zimmermann_2020}. 
Experimental observation of O self-diffusion coefficients vary from 3.35$\times$10$^{-16}$ cm$^2$/s at 300 K \cite{JeongACS2022} to 5.8$\times$10$^{-13}$ cm$^2$/s at 1500 K \cite{Uhlendorf2021, Uhlendorf2023, Uhlendorf2024}.
Moreover, in many wide band gap metal oxides including $\beta$-Ga$_2$O$_3$ \cite{Kohan2000, Peelaers2019}, the Fermi energy  ($\text{E}_\text{Fermi}$) frequently appears close to the conduction band minimum (CBM). 
This high $\text{E}_\text{Fermi}$, associated with intentional or unintentional donor impurities, reduces the formation energies of negatively charged $\text{O}_{\text{i}}$, promoting the role of O$_\text{i}$ in O diffusion \cite{0_Lee_2023}.

$\text{O}_{\text{i}}$ diffusion is also practically important for tuning properties of semiconducting oxides. 
For example, in ZnO and TiO$_2$, the injection and subsequent diffusion of $\text{O}_\text{i}$ can help suppress unintentional $\text{V}_\text{O}$ acting as unwanted donors that degrade carrier mobilities \cite{Alberi2016, Zunger2001, Moses2016}. 
Although conventional doping methods are often limited by high-energy processing (e.g., ion implantation) or inadequate control (in-diffusion), recent studies have revealed that surface treatment can dramatically lower the kinetic barriers for $\text{O}_{\text{i}}$ incorporation and formation. 
Injecting O from a clean, poison-free surface introduces $\text{O}_\text{i}$ that propagates into the near-surface bulk, effectively annihilating $\text{V}_\text{O}$ at previously unattainable low temperatures \cite{JeongACS2022, JeongJVSTA2023}, credited to notably low diffusion activation energies.  

\begin{figure*}[htbp!]
\centering
\includegraphics[width=1\linewidth]{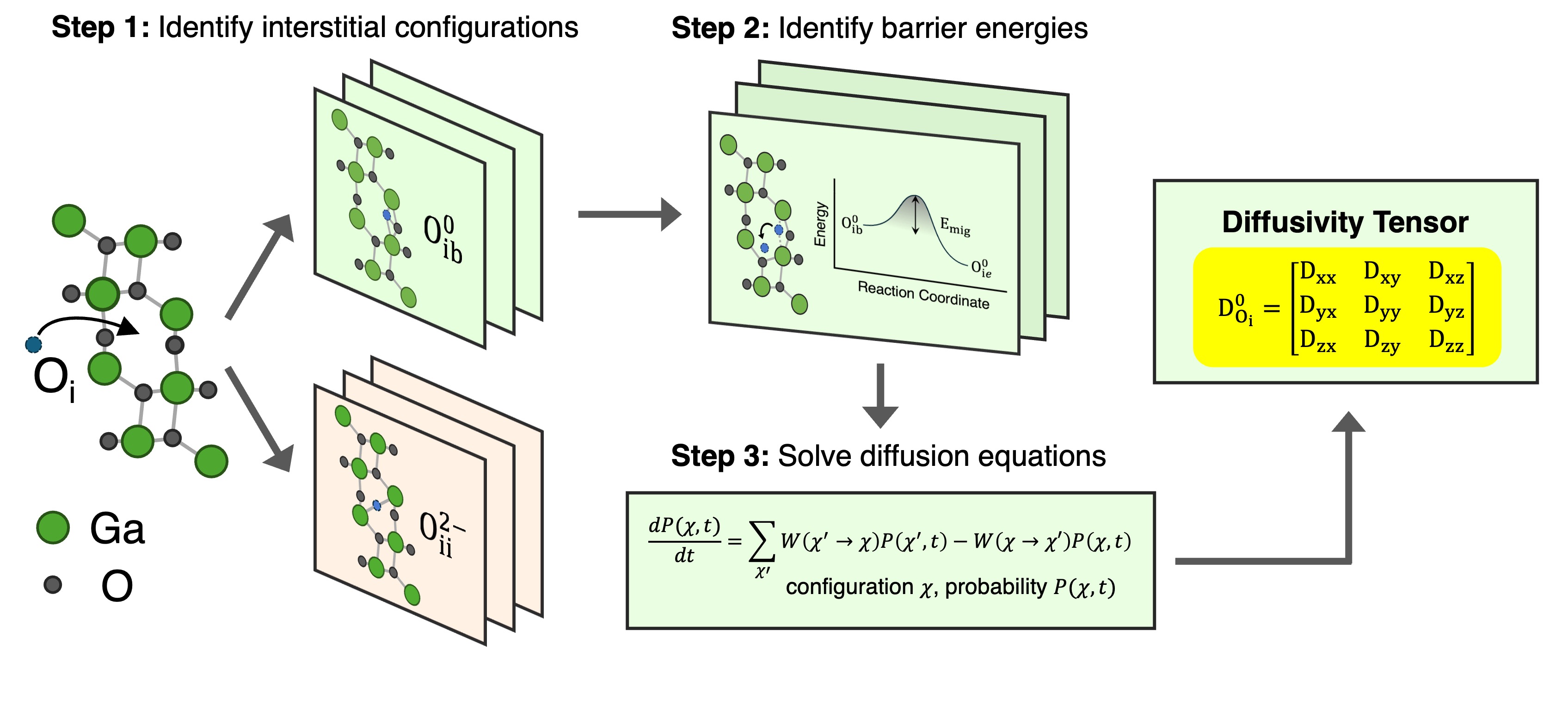}
\caption{Schematic workflow used to construct the three-dimensional diffusivity tensor adapted from Lee \textit{et al.} \cite{CLee2024}.
}
\label{fig:M-workflow}
\end{figure*} 

In this study, we employ a direct approach combining first-principles calculations and the solution of the master diffusion equations to elucidate the full diffusion tensor for neutral and 2- charged $\text{O}_{\text{i}}$ in $\beta$-Ga$_2$O$_3$.
We begin by systematically examining a wide range of $\text{O}_{\text{i}}$ configurations, including split-interstitials, and determine their formation energies. 
We identify six $\text{O}_{\text{i}}^{0}$ and five $\text{O}_{\text{i}}^{2-}$ low-energy configurations.
Based on these configurations, we construct three-dimensional diffusion networks for each charge state, identifying 17 unique hops for $\text{O}_{\text{i}}^{0}$ and 28 for $\text{O}_{\text{i}}^{2-}$.
Using defect formation energies (E$_\text{form}$) and migration barriers (E$_\text{mig}$), we assemble and solve the master diffusion equations, yielding three-dimensional diffusion tensors.
Our results indicate that both $\text{O}_{\text{i}}^0$ and $\text{O}_{\text{i}}^{2-}$ diffuse significantly faster along the $b$-axis, with diffusivities orders of magnitude higher than along the $a^*$- or $c$-axes.
Comparisons to available \ce{^{18}O} tracer self-diffusion experiments demonstrate good agreement with our predictions. 
Finally, because of their lower diffusion barriers and formation energies, we conclude that $\text{O}_{\text{i}}$, opposed to $\text{V}{_{\text{O}}}$, is the likely dominant mediator of O transport in $\beta$-Ga$_2$O$_3$.
Our findings deepen understandings of intrinsic O diffusion in $\beta$-Ga$_2$O$_3$, offering insights to guide future optimization of electronic devices.

\section{Methods}
To predict the $\text{O}_{\text{i}}$ three-dimensional diffusivity tensor, we employed the Onsager approach to construct and solve the master diffusion equations using the package Onsager \cite{20_Trinkle_2016, 21_Trinkle_2017, 22_Trinkle_Onsager}. 
The master diffusion equations are a set of coupled rate equations that describe transitions between different $\text{O}_{\text{i}}$ configurations. 
Their solution results in the Onsager coefficients, i.e.\, the components of the diffusion tensor.
Several simplifying assumptions are embedded in the master diffusion equation approach. 
We treat diffusion as a process in which defect configurations, undergoing harmonic motions, are described by a Markov process with well-defined beginning and end states that thermalize in between jumps. 
The defect concentrations are assumed to follow the grand canonical ensemble, where all species are exchangeable with reservoirs of fixed chemical potential, and site probabilities are governed by the Boltzmann distribution. 
Transition rates are determined by transition state theory, based on the obtained hopping barriers and site energies from one configuration to another.

These rates satisfy detailed balance in equilibrium.
We define $ P(\chi, t) $ as the probability of finding the system in state $ \chi $ at time $ t $, and express the rate of change of the probability as  
\small{
\begin{equation}
    \frac{dP(\chi,t)}{dt} = \sum_{\chi'}\left(W(\chi' \rightarrow \chi) P(\chi',t)-W(\chi \rightarrow \chi') P(\chi,t)\right). \hspace{0.5em}  
\end{equation}
}
Here, for a given configuration $\chi$, the sum is taken over all other possible configurations $\chi'$. 
The first term inside the summation, $ W(\chi' \rightarrow \chi) P(\chi', t) $, represents the rate at which the system transitions from other states $ \chi' $ into state $ \chi $, weighted by the probability of state $ \chi' $. 
The second term, $ W(\chi \rightarrow \chi') P(\chi, t) $, similarly accounts for the rate at which the system leaves state $ \chi $ to transition into other states $ \chi' $.
By incorporating the site energies and migration barriers, we formulate a set of coupled rate equations.
The solution of system of equations under equilibrium ($dP(\chi,t)/dt = 0$, and detailed balance) yields the three-dimensional diffusion tensor for each interstitial network. 
The diagonal elements of the tensor give the diffusion coefficients along the $a^*$, $b$, and $c$ crystallographic directions, while the off-diagonal terms are the cross-diffusion coefficients. 
For additional mathematical and conceptual detail, refer to the original works describing the Onsager software package \cite{20_Trinkle_2016, 21_Trinkle_2017, 22_Trinkle_Onsager, 22.5_Trinkle_Onsager}. 
The overall workflow is illustrated in Figure~\ref{fig:M-workflow}.

In the first phase of our approach (Figure~\ref{fig:M-workflow}, Step 1), we compile a set of candidate $\text{O}_{\text{i}}$ configurations using Voronoi tessellations \cite{36_Goyal_2017}, which partition the lattice into distinct regions around each lattice atom with boundaries defined by points equidistant from neighboring lattice sites.
These points serve as natural candidates for interstitial configurations that minimize repulsion between lattice and interstitial sites.
This approach produced fifteen candidate structures, which we further complemented by configurations proposed in Ingebrigtsen \textit{et al.} and Jeong \textit{et al.} \cite{17_Ingebrigtsen_2018, JeongACS2022}
We relax these initial structures in two dominant charge states (neutral and $2-$ charged) \cite{Jewel2023, Huang2023} using first-principles density functional theory (DFT) \cite{Hohenberg1964, Kohn1965} simulations to obtain site energies. We used the projector augmented wave (PAW) method \cite{25_Bloch_1994, 26_Kresse_1999}, as implemented in the Vienna Ab Initio Simulation Package (VASP) \cite{Kresse1996a, Kresse1996}, and the Perdew-Burke-Ernzerhof (PBE) approximation of the exchange-correlation functional \cite{29_PBE_1996}. 
Calculations employed a plane-wave basis set with a cutoff energy of 420 eV. 
The ground state lattice parameters of the monoclinic $\beta$-Ga$_2$O$_3$ conventional unit cell were determined to be $a =$ 12.28 {\AA}, $b =3.05$ {\AA}, $c = 5.82$ {\AA}, and $\beta = 103.76^\circ$ for 1×4×2 supercell, and are consistent with other computational \cite{13_Kyrtsos_2017,40_Yoshioka_2007,33_Zacherle_2013} and experimental studies \cite{34_Geller_1960,35_Ahman_1996}.
Defect structures were modeled in 160-atom supercells using the Monkhorst-Pack scheme \cite{31_Monkhorst_1976} with a 2$\times$2$\times$2 k-point mesh. 
Geometry optimization of defect structures was performed with a convergence criterion of $1\times10^{-4}$ eV for the total energy and $2\times10^{-2}$ eV/\textup{\AA} for atomic forces.

The defect formation energy $\text{E}_\text{form}[\text{O}_\text{i}^q]$ (site energies) of an interstitial in charge state $q$ is obtained using the supercell approach \cite{37_Freysoldt_2014, 38_Lany_2009, 39_Adamczyk_2021}:
\small{
\begin{equation}
    \text{E}_\text{form}[\text{O}_\text{i}^q] = \text{E}_{\text{tot}}[\text{O}_\text{i}^q] - \text{E}_{\text{tot}}[\text{Bulk}] - \mu_{\text{O}} + q\text{E}_{\text{Fermi}} + \text{E}_\text{corr} \hspace{0.5em} ,
\end{equation}
}
\noindent where $\text{E}_{\text{tot}}[\text{O}_\text{i}^q]$ is the total energy of the defective supercell and $\text{E}_{\text{tot}}[\text{Bulk}]$ is the total energy of the pristine bulk supercell. 
The term $\mu_{\text{O}}$ is the chemical potential of oxygen in the system. 
The charge state is given by $q$, which can take the value of $0$ or $2-$, and $\text{E}_{\text{Fermi}}$ accounts for the exchange of electrons from reference electron reservoir. 
Finally, the energy correction term, $\text{E}_{\text{corr}}$, accounts for the finite-size effects resulting from electrostatic interactions between charged defects in adjacent supercells. 
We utilized the method proposed by Lany and Zunger \cite{38_Lany_2009} to calculate the finite size effect corrections for potential alignment $\Delta \text{E}_{\text{pa}}(D,q)$ and image charge $\Delta \text{E}_{\text{i}}$.
For each charge state, we assemble an independent defect library containing relaxed configurations and their associated $\text{E}_\text{for}$.
To down-select the number of defects included, only configurations with $\text{E}_\text{for}\leq1.2$ eV relative to the lowest energy defect are used to construct the network, resulting in six $\text{O}_\text{i}^{0}$ and five $\text{O}_\text{i}^{2-}$ configurations.

In the second phase (Figure~\ref{fig:M-workflow}, Step 2), we construct the hopping network by considering all possible transitions between interstitial pairs in the supercell, regardless of proximity. 
We saturated the unit cell of $\beta$-Ga$_2$O$_3$ with defects in our library and identify all possible symmetry-unique hops between pairs within 4 {\AA} to populate the network. 
Since many of these hops can be decomposed into sequences of shorter, substituent hops, we use a down-selection approach to isolate independent hops, as discussed in Section III(B). 
We calculated migration barriers for each possible path using the climbing-image nudged elastic band method (ci-NEB) \cite{Henkelman2000}.
At this stage, the three-dimensional diffusion network is complete, providing all necessary inputs for the Onsager methodology. 

\begin{figure*}[htbp!]
\centering
\includegraphics[width=1\linewidth]{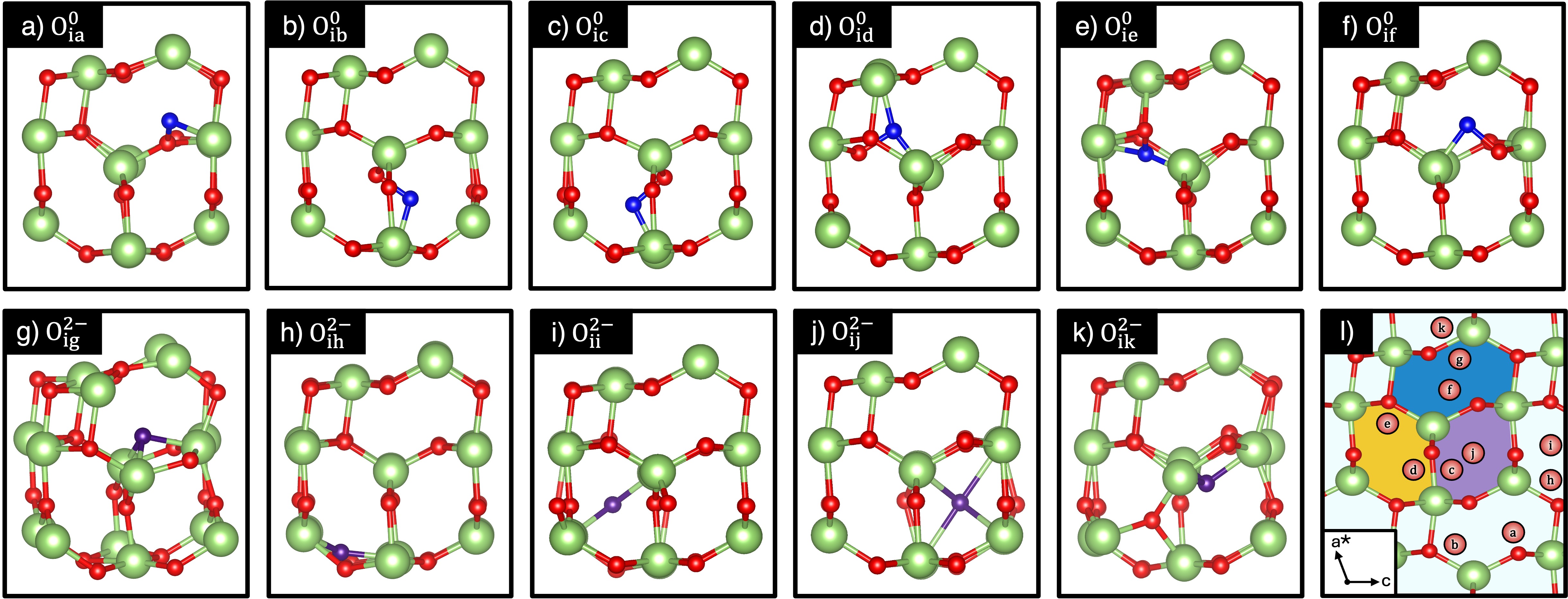}
\caption{Low-energy configurations of (a-f) $\text{O}_\text{i}^{0}$ and (g-k) $\text{O}_\text{i}^{2-}$. Schematic representation (l) of $b$-axis channels in $\beta$-Ga$_2$O$_3$, conventionally named "A", "B", and "C", depicted in blue, gold, and purple, respectively, with labeled positions of unique $\text{O}_\text{i}$ sites.
}
\label{fig:R-neutral_charged_lib}
\end{figure*}

In the third, final phase of the workflow (Figure~\ref{fig:M-workflow}, Step 3), we assemble and solve the master diffusion equations.
These equations are constructed using the interstitial sites and their site energies, along with the transitions between them and associated energy barriers, comprising the hopping network.
The master diffusion equations are derived from the Onsager reciprocal relations, which describe a generalized linear relationship between thermodynamic forces and fluxes.
The constants of proportionality in this relationship represent diffusion coefficients. 
Details of the formulations, assumptions of the Onsager approach, and parameters used in the DFT simulations to compute site and transition energies are provided in the SI.

\section{Results and Discussion}
\subsection{Oxygen Interstitial Defect Configurations and Formation Energies}
In highly anisotropic, monoclinic $\beta$-Ga$_2$O$_3$, local Ga-O bonding environments vary extensively, resulting in a broad spectrum of extended defects, including split configurations.
Previous studies of Ga interstitials ($\text{Ga}_{\text{i}}$) and vacancies ($\text{V}_{\text{Ga}}$), for instance, have revealed that two distinct Ga sites (octahedral and tetrahedral) can adopt extended “N-split” configurations spanning multiple sites (e.g., $\text{Ga}_{\text{i}}$–$\text{V}_{\text{Ga}}$–$\text{Ga}_{\text{i}}$ or $\text{V}_{\text{Ga}}$–$\text{Ga}_{\text{i}}$–$\text{V}_{\text{Ga}}$ chains), which can be more stable than conventional point defects \cite{10_Frodason_2023,CLee2024}.
Similarly, three inequivalent O sites, including two 3-fold configurations (trigonal planar coordination) and one 4-fold arrangement (distorted tetrahedron) can adopt multiple configurations, highlighting the need for a comprehensive examination of possible split $\text{O}_{\text{i}}$.
In this work, we have identified total 11 symmetry-unique defect configurations, including both conventional and split $\text{O}_{\text{i}}$.

Figure~\ref{fig:R-neutral_charged_lib}(a-f) features six low-energy configurations of $\text{O}_\text{i}^{0}$, while Figure~\ref{fig:R-neutral_charged_lib}(g-k) features five $\text{O}_\text{i}^{2-}$ configurations, both of which are used to construct their respective defect libraries.
These libraries include both split-interstitials (a-f,k), where the defective O shares a site with a lattice O atom, and traditional isolated interstitials (g-j) where the defective O sits in a channel with minimal displacement of lattice O atoms.

\begin{figure}[!hbtp]
\centering
\includegraphics[width=1\linewidth]{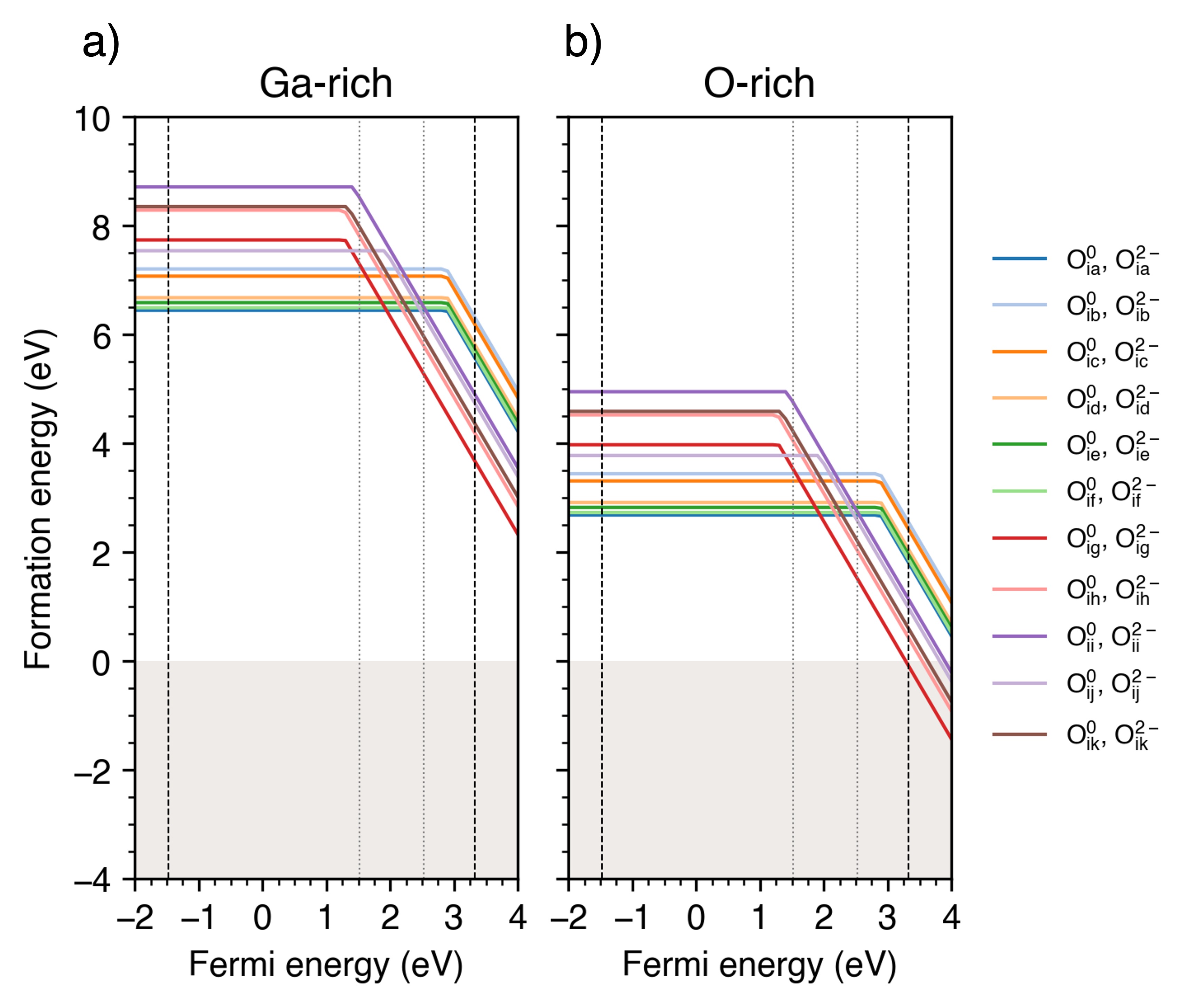}
\caption{Formation energies of various configurations of $\text{O}_{\text{i}}$ as a function of Fermi-energy level using the PBE level of theory under (a) Ga-rich and (b) O-rich thermodynamic conditions. The dashed lines near the left and right sides of the plot are the HSE valence band maximum (VBM) and conduction band minimum (CBM) levels, respectively, predicted by band alignments using the electrostatic potentials between PBE and HSE band structures. The leftmost dotted line is the Fermi energy chosen to represent the neutral defects in the self-diffusivity plots (1.8 eV below CBM) and the rightmost dotted line is the Fermi energy chosen to represent the charged defects in the self-diffusivity plots (0.8 eV below CBM).  }
\label{fig:R-DFE}
\end{figure}

Among neutral interstitials, $\text{O}_{\text{ia}}^0$ exhibits the lowest formation energy of 2.68 eV (Figure~\ref{fig:R-DFE}), followed by $\text{O}_{\text{if}}^0$, $\text{O}_{\text{ie}}^0$, and $\text{O}_{\text{id}}^0$, which have energies $0.05$, $0.14$, and $0.23$ eV higher than $\text{O}_{\text{ia}}^0$, respectively.
These four neutral defects are all within the A channel (see Figure~\ref{fig:R-neutral_charged_lib}(l)), consistent with the trend observed in Blanco \textit{et al.} \cite{14_Blanco_2005}.
The remaining two defects, $\text{O}_{\text{ib}}^0$ and $\text{O}_{\text{ic}}^0$, within the B and C channels, are $0.76$ and $0.62$ eV higher than $\text{O}_{\text{ia}}^0$, respectively.
Other studies have also identified the split-interstitial $\text{O}_{\text{if}}$ \cite{Peelaers2019, Lehtom2020, 17_Ingebrigtsen_2018, Tuttle2023, Deak2017, Sun2019}, as well as $\text{O}_{\text{ic}}$\cite{Li2024} and $\text{O}_{\text{ie}}$ \cite{Tuttle2023}.

\begin{figure*}[htbp!]
\centering
\includegraphics[width=1\linewidth]{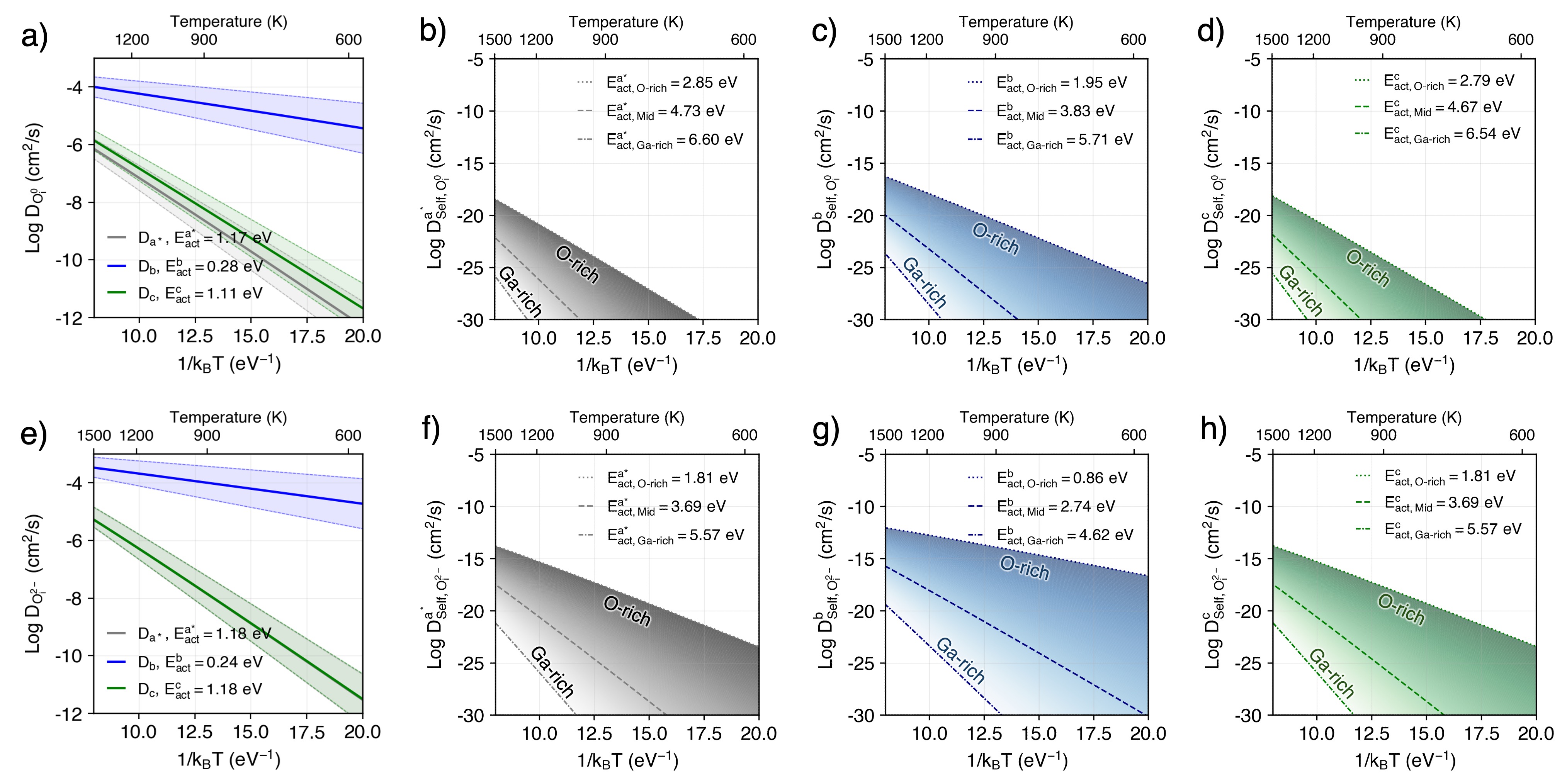}
\caption{
Arrhenius plot of diffusivities of (a) $\text{O}_{\text{i}}^0$ and (e) $\text{O}_{\text{i}}^{2-}$. Shaded regions in (a) and (b) represent 0.1 eV uncertainty in E$_\text{mig}$. The remaining plots highlight self-diffusivities of (b, c, d) $\text{O}_{\text{i}}^0$ and (f, g, h) $\text{O}_{\text{i}}^0$. Thermodynamic limits (Ga-rich and O-rich) are given within the dotted lines, where the dashed middle line indicates the midpoint between two limits. Activations energies (E$_\text{act}$) for various directions (a,e) and thermodynamic conditions (a-c,f-h) are indicated. 
}
\label{fig:R-diff_self-diff}
\end{figure*} 

For charged interstitials, the defect in the A channel, $\text{O}_{\text{ig}}^{2-}$, exhibits the lowest E$_\text{form}$ of 1.56 eV when E$_\text{Fermi}$ is at 0.8 eV below the CBM.
Two configurations close to the lattice in the B and C channels, $\text{O}_{\text{ih}}^{2-}$ and $\text{O}_{\text{ik}}^{2-}$, have E$_\text{form}$ higher than that of $\text{O}_{\text{ig}}^{2-}$ by $0.52$ and $0.69$ eV, respectively.
In contrast, $\text{O}_{\text{ii}}^{2-}$ and $\text{O}_{\text{ij}}^{2-}$, both in the centers of the B and C channels, exhibit the highest E$_\text{form}$, higher that of $\text{O}_{\text{ig}}^{2-}$ by $1.06$ and $1.23$ eV, respectively.

Electrostatic interactions between lattice atoms and $\text{O}{\text{i}}$ likely drive this trend in E$_\text{form}[\text{O}_{\text{i}}^{2-}]$. 
For instance, $\text{O}_{\text{ig}}^{2-}$ sits father from lattice O atoms within the A channel to reduce repulsive interactions, while simultaneously maintaining optimal distance to nearby Ga atoms to maximize electrostatic attractions.
All other defects are relatively closer to lattice O atoms, farther away from Ga atoms, or both, consistent with the highest energy defects in the library. 
Other studies \cite{18_Zimmermann_2020, Lehtom2020, Li2024, 17_Ingebrigtsen_2018, 14_Blanco_2005, 33_Zacherle_2013, Martins2024, JeongACS2022} have reported two additional configurations in the center of the A channel.
Geometry optimization reveals that the 2-fold coordinated configurations reported in Zimmermann \textit{et al.} and Li \textit{et al.} relax to $\text{O}_{\text{ig'}}$ (Figure S1(b)).
While the 3-fold coordinated configurations $\text{O}_{\text{ig'}}$ and $\text{O}_{\text{ig''}}$ (Figure S1(c)) reported in Zimmermann \textit{et al.}, Lehtom \textit{et al.}, and Jeong \textit{et al.} were confirmed stable, they were excluded from the diffusion network following a detailed analysis of diffusion pathways, as described in Section III(B).
Overall, our predicted formation energies of neutral and charged $\text{O}_{\text{i}}$ are in good agreement with those reported in Zacherele \textit{et al.} and Peelars \textit{et al.} \cite{33_Zacherle_2013, Peelaers2019}.

\subsection{Diffusion Networks and Migration Activation Energies}   

To create diffusivity tensors, we identify and construct core sets of ``principal hops'' (PHs), the fundamental and indivisible migrations, for both the neutral and charged diffusion networks.
In other words, a PH is one that cannot be decomposed into a sequence of smaller transitions. 
To identify PHs, we enumerate all transitions between defect pairs across all possible sites of the supercell. 
For split-interstitial defects, we use a representative midpoint between two oxygen atoms to denote the defect site and efficiently track interstitialcy (kick-out) hops. 
In these hops, the original interstitial atom occupies the lattice site and the original lattice atom migrates as an interstitial.

Enumerating each possible migration from a defect-saturated supercell yields hundreds of pathways.
To narrow these to a practical number, we remove symmetry-equivalent hops and systematically eliminate candidates that could be broken down into smaller PHs.
For example, instead of a direct hop from $\text{O}_{\text{ia}}^{0}$ to $\text{O}_{\text{ic}}^{0}$, inspecting intermediate configurations between them suggests that $\text{O}_{\text{i}}$ migrates first through $\text{O}_{\text{ib}}^{0}$, resulting in two PHs.
We perform initial coarse NEB calculations to identify additional metastable states along selected paths and eliminate unrealistic paths with barriers exceeding 10 eV.
When metastable intermediate configurations emerge, we conduct geometry relaxations to determine if they are within our network or new candidate defects. 
In all cases, these configurations match or closely resemble existing interstitials in our libraries. 
If migration barriers exceed 10 eV, we eliminate the hop.
Through these down-selection processes, we reduce the number of hops in the neutral network from 125 to 17, and from 140 to 28 in the charged network. 
Once a single pathway cannot be decomposed into shorter hops, we perform a final NEB calculation with more stringent convergence criteria to identify the migration barrier, E$_\text{mig}$.
These identified symmetry-unique PHs and corresponding E$_\text{mig}$'s for $\text{O}_{\text{i}}^0$ and $\text{O}_{\text{i}}^{2-}$ are summarized in Tables S1 and S2, respectively. 
We then used the Onsager software package to find diffusivities of $\text{O}_{\text{i}}^0$ and $\text{O}_{\text{i}}^{2-}$  (D$_{\text{O}_{\text{i}}^{0}}$ and D$_{\text{O}_{\text{i}}^{2-}}$) \cite{22_Trinkle_Onsager}.

\begin{figure*}[htbp!]
    \centering
    \includegraphics[width=1\linewidth]{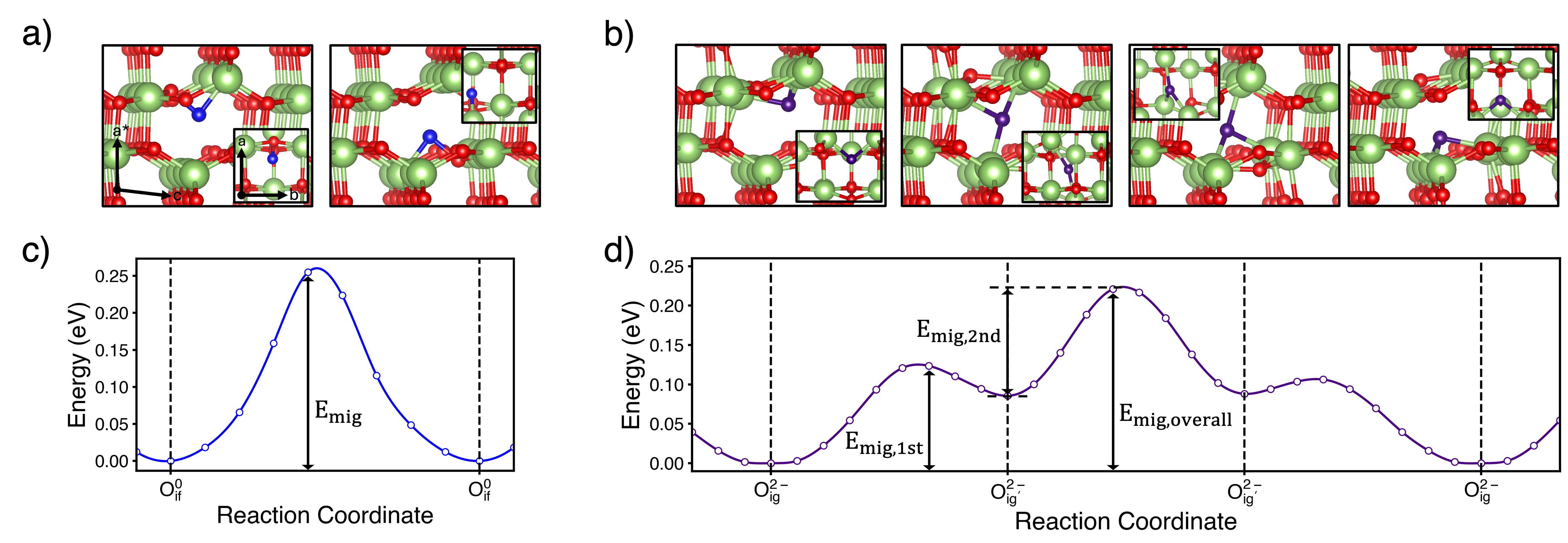}
    \caption{Dominant $b$-axis diffusion pathway of (a) $\text{O}_{\text{if}}^{0}$ and (b) $\text{O}_{\text{ig}}^{2-}$ along with corresponding energy landscapes (c) and (d), respectively, where the energy maxima correspond to 0.25 and 0.24 eV. Dashed lines on the reaction coordinate indicate stable or intermediate configurations.}
    \label{fig:R-b_pathway}
\end{figure*}

Figure~\ref{fig:R-diff_self-diff}(a,e) summarizes predicted diffusivities (D(T)) along the $a^*$, $b$, and $c$ crystallographic directions, following the Arrhenius relationship below:
{\small
\begin{equation}
   \text{D(T)} = \text{D}_0 \exp\left(-\frac{\text{E}_\text{act}}{k_B T}\right),
\end{equation}}

where D$_0$ is the pre-exponential factor and $k_B$ is the Boltzmann constant. 
The shaded region indicates the incorporation of a  $\pm 0.1$ eV uncertainty on the calculated NEB barriers, to account for typical imperfections in DFT simulations. 
Both charge states possess sizable anisotropy, with $b$-axis diffusivity being several orders of magnitude larger than that along the $a^*$ and $c$-axes.
This contrasts with Ga diffusion in $\beta\text{-Ga}_2\text{O}_3$, where diffusion along the $c$-axis dominates for both interstitials and vacancies \cite{CLee2024, 10_Frodason_2023}.
Effective $\text{E}_\text{act}$ for $b$-axis diffusion are $0.28$ eV for the neutral network and $0.24$ eV for the charged network, while those of the $a^*$ and $c$-axes both exceed $1$ eV. 
In the neutral network, $\text{E}_\text{act}^{a^*} = 1.17$ eV and $\text{E}_\text{act}^{c} = 1.11$ eV, while $\text{E}_\text{act}^{a^*} = \text{E}_\text{act}^{c} = 1.18$ eV in the charged network.

Next, the $\text{O}_{\text{i}}$ self-diffusivities (D$_{\text{Self}, \text{O}_\text{i}}$) are estimated from predicted D$_{\text{O}_\text{i}}$.
While D$_{\text{O}_\text{i}}$ presented above represent the diffusivity of isolated interstitials, they differ from values typically measured in experiment, where the concentration of the diffusing species—and consequently the defect E$_\text{form}$—are encompassed.
The self-diffusivities are given by 
{\small
\begin{equation}
\text{D}_{\text{Self}, \text{O}_{\text{i}}} = \frac{[\text{O}_{\text{i}}]}{[\text{O}_{\text{total}}]} \text{D}_{\text{O}_{\text{i}}} \hspace{0.5em}, 
\end{equation}}
where $[\text{O}_{\text{i}}]$ is the concentration of migrating O interstitials, $[\text{O}_{\text{total}}]$ is the concentration of O sites within the lattice, and D$_{\text{O}_{\text{i}}}$ is the diffusivity given in Figure~\ref{fig:R-diff_self-diff}(a,e).
Detailed chemical potential boundaries for O are described in SI Section 4.
Figure~\ref{fig:R-diff_self-diff}(b-d,f-h) depicts Arrhenius plots for $\text{O}_{\text{i}}^0$ and $\text{O}_{\text{i}}^{2-}$ self-diffusivities for a range of thermodynamic conditions. 
We choose a single $\text{E}_\text{Fermi}$ of 3.0 eV above the VBM (1.8 eV below the conduction band minimum (CBM)) for ease of representation of the neutral (Figure~\ref{fig:R-diff_self-diff}(b-d)) and $\text{E}_\text{Fermi}=3.8$ eV (0.8 eV below CBM) for the charged plots (Figure~\ref{fig:R-diff_self-diff}(f-h)). 
The different $\text{E}_\text{Fermi}$ correspond to values for which the neutral and ionized interstitial are favorable. 
Overall, D$_{\text{Self}, \text{O}_\text{i}}$ exhibits similar trends to D$_{\text{O}_\text{i}}$: rapid $b$-axis diffusion dominates, while smaller diffusivities with comparable magnitudes are observed along the $a^*$- and $c$-axes.
Under O-rich conditions, predicted D$_{\text{Self},\text{O}_\text{i}}^{\text{b}}$ at 1200 K are 2.34$\times$10$^{-18}$ and 2.37$\times$10$^{-13}$ for the neutral and charged $\text{O}_{\text{i}}$, respectively.
Generally, D$_{\text{Self}, \text{O}_\text{i}^{2-}}$ are substantially greater than D$_{\text{Self}, \text{O}_\text{i}^0}$ due to lower E$_\text{form}$ of the configurations in the charged library, resulting in higher defect concentrations within the crystal.

By systematically removing individual PHs and recalculating diffusivities, we identify the dominant hopping pathways in each crystallographic direction.
The dominant diffusion pathways along the fast $b$-axis for $\text{O}_{\text{i}}^0$ and $\text{O}_{\text{i}}^{2-}$ are given in Figure~\ref{fig:R-b_pathway}(a,b) with corresponding energy profiles along the minimum energy pathway (MEP) in Figure~\ref{fig:R-b_pathway}(c,d). 
The dominant neutral pathway (PH 17 in Table S1) depicted in Figure~\ref{fig:R-b_pathway}(a) features the migration of $\text{O}_{\text{if}}^0$ along distinct, well-defined trajectories within the A channel. 
The migration barrier at the peak of Figure~\ref{fig:R-b_pathway}(c) is 0.25 eV, enabling remarkably rapid diffusion along the $b$-axis.
Several other PHs between $\text{O}_{\text{ia}}^0$ and $\text{O}_{\text{if}}^0$ create low-barrier pathways along the $b$-axis (Figure S2(c,d)), but with slightly larger E$_\text{mig}$ (Figure S2(e)).

The dominant charged pathway (PH 18 in Table S2), depicted in Figure~\ref{fig:R-b_pathway}(b), similarly traverses the A channel, but unlike PH 17, it includes an intermediate, metastable state $\text{O}_{\text{ig'}}$ (Figure S1(b)).
Migration along the MEP in Figure~\ref{fig:R-b_pathway}(d) begins with the lowest-energy configuration, $\text{O}_{\text{ig}}^{2-}$, transitions over the first energy barrier (E$_\text{mig, 1st}$ = 0.12 eV) and proceeds through the intermediate state, $\text{O}_{\text{ig'}}^{2-}$.
It then crosses another energy barrier (E$_\text{mig, 2nd}$ = 0.14 eV) and passes to the next equivalent $\text{O}_{\text{ig'}}^{2-}$ and $\text{O}_{\text{ig}}^{2-}$ states.
This pathway exhibits a slightly smaller overall energy maxima (E$_\text{mig, overall}$ = 0.24 eV) compared to that of the neutral path, enabling exceptionally fast $b$-axis diffusion.
The formation and breakage of bonds with neighboring Ga atoms likely facilitates the stabilization of $\text{O}_{\text{ig'}}^{2-}$.
Additionally, the 2- charge state could induce higher electrostatic attractions with lattice Ga atoms, further stabilizing the intermediate configurations and slightly lowering the overall barrier.
Despite its stability, we exclude $\text{O}_{\text{ig’}}$ from our diffusion network because its bonding environment and E$_\text{form}$ closely resemble those of $\text{O}_{\text{ig}}$ (Figure S1(a,b)).
While their states are distinct at 0 K, we expect $\text{O}_{\text{ig}}$ and $\text{O}_{\text{ig’}}$ to become nearly identical at slightly elevated T due to additional thermal vibrations.
The MEP closely resembles that identified in Zimmermann \textit{et al.}, with intermediate configurations similar to $\text{O}_{\text{ig’’}}$ (Figure S1(c)) and distinct end states \cite{18_Zimmermann_2020}.
The migration barrier E$_\text{mig, 2nd}=0.14$ for PH 18 matches the E$_\text{mig}$ reported in Zimmermann \textit{et al.}, potentially suggesting similar local environments or structural constraints shaping the energy landscape.

\begin{figure}
    \centering
    \includegraphics[width=1\linewidth]{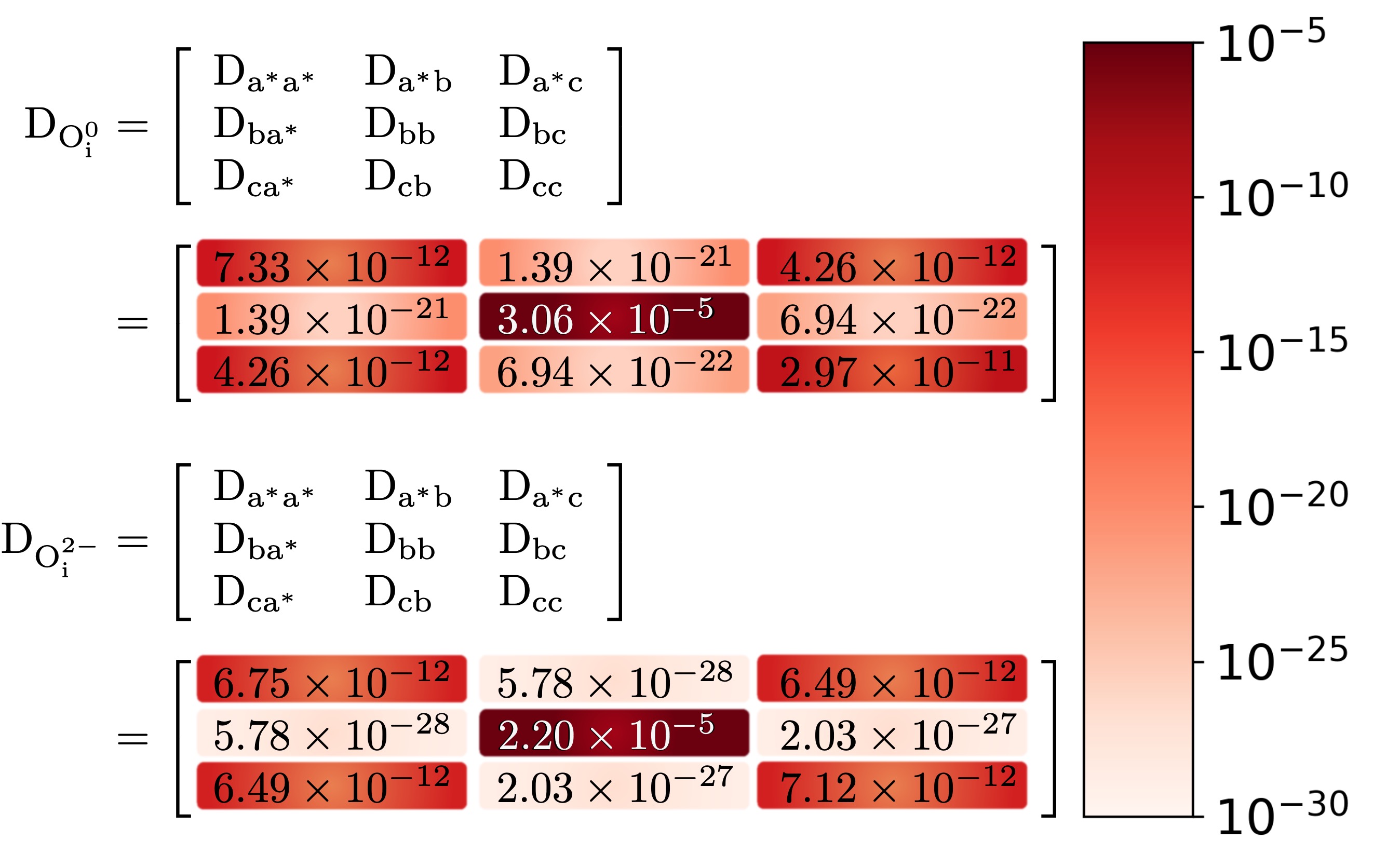}
    \caption{Three-dimensional diffusivity tensors of $\text{O}_{\text{i}}^0$ and $\text{O}_{\text{i}}^{2-}$ at 600 K ($\text{cm}^2/\text{s}$).}
    \label{fig:R-600K_tensor}
\end{figure}

While significantly smaller, $a^*$- and $c$-axis diffusion exhibit interesting behavior.
As revealed in Figure~\ref{fig:R-diff_self-diff}(a,e), these directions exhibit E$_\text{act}$ and D$_{\text{O}_\text{i}}$ that are very similar to each other, not observed in either $\text{V}_\text{Ga}$ or $\text{Ga}_\text{i}$ diffusion \cite{CLee2024}. 
In the charged network, D$_{ca^*}$ is nearly identical to its pure component counterparts, D$_{a^*a^*}$ and D$_{cc}$, even at lower temperatures where these differences are among the largest (Figure~\ref{fig:R-600K_tensor}).
This suggests that diffusion along the $a^*$ and $c$-axes is coupled for $\text{O}_{\text{i}}^{2-}$.
The ``dominant hop'' analysis reveals that the same set of principal hops (PH 20, 32, 33, 40 in Table S2) contribute to both $a^*$ and $c$-axis diffusion.
Removing any of these hops reduces D$_{a^*a^*}$ and D$_{cc}$ by up to 90\%.
The PHs 20, 32, and 33 form a fully connected dominant pathway for $a^*/c$-axis diffusion, as detailed in Figure S2.
In the neutral network, D$_{ca^*}$ is smaller than both D$_{a^*a^*}$ and D$_{cc}$, but remains closer in magnitude to D$_{a^*a^*}$ than to D$_{cc}$.
Removing each dominant $a^*$-axis hop reduces diffusion exclusively along the $a^*$-direction, whereas removing dominant $c$-axis hops substantially decreases diffusion along both $a^*$ and $c$-directions.
This result suggests that diffusion of $\text{O}_{\text{i}}^{0}$ along the $a^*$-axis is dependent on some $c$-axis hops, while $c$-axis diffusion can exist on its own.
The main $c$-axis pathway and possible $a$-axis pathways are illustrated in Figures S3 and S4, respectively, where the $a$-axis pathways highlighted in orange indicate dependence on a $c$-axis hop. 

\subsection{Comparison to Experiments and Oxygen Vacancy Diffusion}

Finally, the self-diffusivities in Figure~\ref{fig:R-diff_self-diff}(f-h) are compared to previously reported experimental results. 

\begin{figure}[htbp!]
    \centering
    \includegraphics[width=1\linewidth]{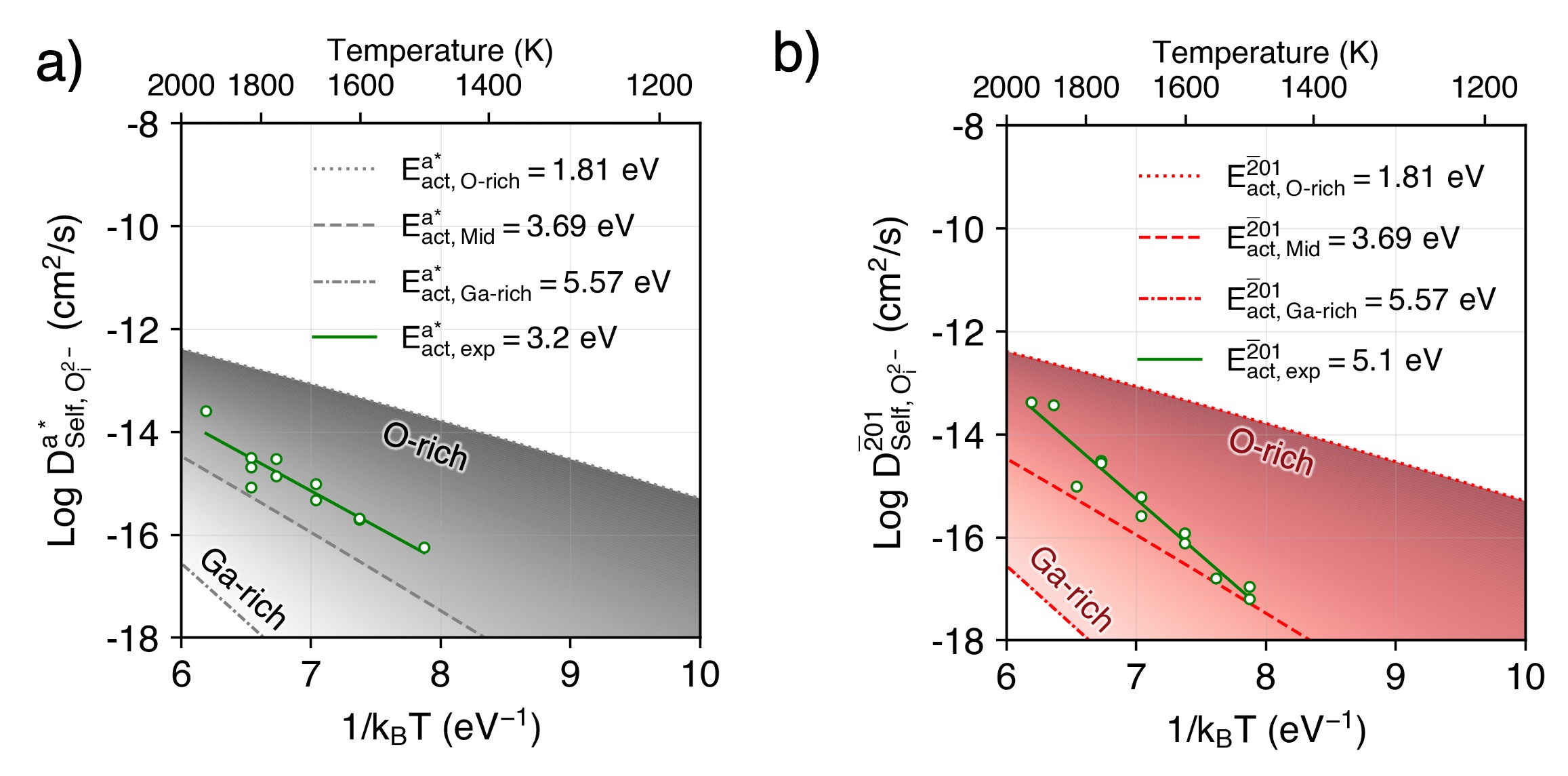}
    \caption{Arrhenius plots of self-diffusivities with experimental data collected along (a) [100] (Ref. [\onlinecite{Uhlendorf2021}]) and (b) [$\overline{2}01$] (Ref. [\onlinecite{Uhlendorf2023}]) from \ce{^{18}O} tracer studies. The thermodynamic limits (Ga-rich and O-rich) are given within the dotted lines, where the dashed middle line indicates the midpoint between two limits. Activations energies (E$_\text{act}$) for various thermodynamic conditions are indicated.}
    \label{fig:R-experiment}
\end{figure} 

\noindent Figure~\ref{fig:R-experiment} presents experimentally measured D$_{\text{Self}, \text{O}_{\text{i}}}$ and $\text{E}_\text{act}$ from Uhlendorf \textit{et al.}, obtained from \ce{^{18}O} tracer experiments conducted at temperatures between 1400 and 2000 K in unintentionally doped (UID) $\beta$-Ga$_2$O$_3$ along (a) [100] \cite{Uhlendorf2021} and (b) [$\overline{2}01$] \cite{Uhlendorf2023}.
The measured D$_{\text{Self}, \text{O}_{\text{i}}}$ fall within the range of our prediction for O$_\text{i}^{2-}$ within the slightly O-rich region, consistent with our expectation given the experimental setup \cite{Uhlendorf2021, Uhlendorf2023, Uhlendorf2024}.
Our predicted $\text{E}_\text{act}$ ranges from $1.81$ (O-rich) to $5.57$ eV (Ga-rich) along [100] and [$\overline{2}01$], when E$_\text{Fermi}$ is 3.8 eV.
The experimentally measured $\text{E}_\text{act}^{100}$ of $3.2 \pm 0.4$ and $\text{E}_\text{act}^{\overline{2}01}$ of $5.0 \pm 0.4$ eV lie in the predicted range of our predictions \cite{Uhlendorf2021, Uhlendorf2023}.
It is worth noting that these predicted ranges are sensitive to the position of E$_\text{Fermi}$, which can make direct comparisons with experimental data difficult.
Nevertheless, the overall consistency between our findings and experimental measurements on D$_{\text{Self}, \text{O}_{\text{i}}}$ and $\text{E}_\text{act}$ highlights the robustness and reliability of our approach.

We also compare our $\text{E}_\text{act}$ for $\text{O}_{\text{i}}$ self-diffusion with estimated $\text{E}_\text{act}$ values for V$_\text{O}$ self-diffusion, derived from previously reported $\text{E}_\text{for}$ and $\text{E}_\text{mig}$ from first-principles calculations, to assess whether O self-diffusion in $\beta$-Ga$_2$O$_3$ is most dominantly interstitial- or vacancy-mediated.
We adopt the range of $\text{E}_\text{for}$ values from Kyrtsos \text{et al.} (3.8$-$4.6 eV) under O-rich and n-doped conditions, along with $\text{E}_{\text{mig}}$ (1.2$-$2.7 eV) \cite{13_Kyrtsos_2017}.
By adding these $\text{E}_{\text{for}}$ and $\text{E}_{\text{mig}}$, we estimate E$_\text{act}$ of V$_\text{O}$ self-diffusion between 5.0$-$7.3 eV, which is far beyond our predicted range for O$_\text{i}$ self-diffusion E$_\text{act, O-rich}$ between 0.86$-$2.85 eV (Figure~\ref{fig:R-diff_self-diff}).
Overall, our results are in agreement with Uhlendorf \textit{et al.}'s hypothesis suggesting that O diffusion in $\beta$-Ga$_2$O$_3$ is likely governed by $\text{O}_\text{i}$, not $\text{V}_\text{O}$ \cite{Uhlendorf2021, Uhlendorf2023, Uhlendorf2024}.

\section{Conclusion}
From first-principles calculations, we investigated various oxygen defects in $\beta$-Ga$_2$O$_3$, including split-interstitial configurations, and examined the associated diffusion pathways. 
Employing the Onsager formalism, we developed and solved the master diffusion equations for $\text{O}_{\text{i}}^0$ and $\text{O}_{\text{i}}^{2-}$ under detailed balance, yielding three-dimensional diffusivity tensors.
Our results revealed that the $b$-axis enables the fastest diffusion, with diffusivities of 2.30$\times$10$^{-4}$ $\text{cm}^2/\text{s}$ and 6.33$\times$10$^{-5}$ $\text{cm}^2/\text{s}$ at 1200 K, and corresponding effective activation energies of 0.28 and 0.24 eV for $\text{O}_{\text{i}}^0$ and $\text{O}_{\text{i}}^{2-}$, respectively.
At O-rich conditions, we predicted self-diffusivities of 2.34$\times$10$^{-18}$ and 2.37$\times$10$^{-13}$ $\text{cm}^2/\text{s}$ along the $b$-axis at 1200 K, for $\text{O}_{\text{i}}^0$ and $\text{O}_{\text{i}}^{2-}$, respectively.
We identified the most dominant diffusion pathways along each crystallographic direction and found that O self-diffusion is likely mediated by interstitials, consistent with reported \ce{^{18}O} tracer diffusion experiments. 
Our findings expand our insights into mass transport and defect-mediated degradation mechanisms in $\beta$-Ga$_2$O$_3$, contributing to a broader perspective on defect diffusion in semiconducting oxides. 

\section{Supplementary Material}
\noindent See supporting information for Tables S1 and S2, Figures S1-S5, and details on oxygen chemical potential boundary selection.

\section{Acknowledgment}
\noindent GM and EE acknowledge support from the U.S. National Science Foundation (NSF) via Grant No. 1922758 (DIGI-MAT). CL and EE acknowledge funding provided by the Air Force Office of Scientific Research under Award No. FA9550-21-0078 (Program Manager: Dr. Ali Sayir). 
This work used PSC Bridges-2 at the Pittsburgh Supercomputing Center through allocation MAT220011 from the Advanced Cyberinfrastructure Coordination Ecosystem: Services \& Support (ACCESS) program, which is supported by National Science Foundation grants \#2138259, \#2138286, \#2138307, \#2137603, and \#2138296.

\section{DATA AVAILABILITY}
\noindent The data that supports the findings of this study are available on GitHub at https://github.com/ertekin-research-group/2025-Ga2O3-O-Diffusion. 

\bibliography{01_references}

\begin{thebibliography}{66}%
\makeatletter
\providecommand \@ifxundefined [1]{%
 \@ifx{#1\undefined}
}%
\providecommand \@ifnum [1]{%
 \ifnum #1\expandafter \@firstoftwo
 \else \expandafter \@secondoftwo
 \fi
}%
\providecommand \@ifx [1]{%
 \ifx #1\expandafter \@firstoftwo
 \else \expandafter \@secondoftwo
 \fi
}%
\providecommand \natexlab [1]{#1}%
\providecommand \enquote  [1]{``#1''}%
\providecommand \bibnamefont  [1]{#1}%
\providecommand \bibfnamefont [1]{#1}%
\providecommand \citenamefont [1]{#1}%
\providecommand \href@noop [0]{\@secondoftwo}%
\providecommand \href [0]{\begingroup \@sanitize@url \@href}%
\providecommand \@href[1]{\@@startlink{#1}\@@href}%
\providecommand \@@href[1]{\endgroup#1\@@endlink}%
\providecommand \@sanitize@url [0]{\catcode `\\12\catcode `\$12\catcode `\&12\catcode `\#12\catcode `\^12\catcode `\_12\catcode `\%12\relax}%
\providecommand \@@startlink[1]{}%
\providecommand \@@endlink[0]{}%
\providecommand \url  [0]{\begingroup\@sanitize@url \@url }%
\providecommand \@url [1]{\endgroup\@href {#1}{\urlprefix }}%
\providecommand \urlprefix  [0]{URL }%
\providecommand \Eprint [0]{\href }%
\providecommand \doibase [0]{https://doi.org/}%
\providecommand \selectlanguage [0]{\@gobble}%
\providecommand \bibinfo  [0]{\@secondoftwo}%
\providecommand \bibfield  [0]{\@secondoftwo}%
\providecommand \translation [1]{[#1]}%
\providecommand \BibitemOpen [0]{}%
\providecommand \bibitemStop [0]{}%
\providecommand \bibitemNoStop [0]{.\EOS\space}%
\providecommand \EOS [0]{\spacefactor3000\relax}%
\providecommand \BibitemShut  [1]{\csname bibitem#1\endcsname}%
\let\auto@bib@innerbib\@empty
\bibitem [{\citenamefont {Tippins}(1965)}]{Tippins1965}%
  \BibitemOpen
  \bibfield  {author} {\bibinfo {author} {\bibfnamefont {H.~H.}\ \bibnamefont {Tippins}},\ }\href {https://doi.org/10.1103/PhysRev.140.A316} {\bibfield  {journal} {\bibinfo  {journal} {Phys. Rev.}\ }\textbf {\bibinfo {volume} {140}},\ \bibinfo {pages} {A316} (\bibinfo {year} {1965})}\BibitemShut {NoStop}%
\bibitem [{\citenamefont {Matsumoto}\ \emph {et~al.}(1974)\citenamefont {Matsumoto}, \citenamefont {Aoki}, \citenamefont {Kinoshita},\ and\ \citenamefont {Aono}}]{Matsumoto_1974}%
  \BibitemOpen
  \bibfield  {author} {\bibinfo {author} {\bibfnamefont {T.}~\bibnamefont {Matsumoto}}, \bibinfo {author} {\bibfnamefont {M.}~\bibnamefont {Aoki}}, \bibinfo {author} {\bibfnamefont {A.}~\bibnamefont {Kinoshita}},\ and\ \bibinfo {author} {\bibfnamefont {T.}~\bibnamefont {Aono}},\ }\href {https://doi.org/10.1143/JJAP.13.1578} {\bibfield  {journal} {\bibinfo  {journal} {Japanese Journal of Applied Physics}\ }\textbf {\bibinfo {volume} {13}},\ \bibinfo {pages} {1578} (\bibinfo {year} {1974})}\BibitemShut {NoStop}%
\bibitem [{\citenamefont {Ueda}\ \emph {et~al.}(1997)\citenamefont {Ueda}, \citenamefont {Hosono}, \citenamefont {Waseda},\ and\ \citenamefont {Kawazoe}}]{Ueda1997}%
  \BibitemOpen
  \bibfield  {author} {\bibinfo {author} {\bibfnamefont {N.}~\bibnamefont {Ueda}}, \bibinfo {author} {\bibfnamefont {H.}~\bibnamefont {Hosono}}, \bibinfo {author} {\bibfnamefont {R.}~\bibnamefont {Waseda}},\ and\ \bibinfo {author} {\bibfnamefont {H.}~\bibnamefont {Kawazoe}},\ }\href {https://doi.org/10.1063/1.119693} {\bibfield  {journal} {\bibinfo  {journal} {Applied Physics Letters}\ }\textbf {\bibinfo {volume} {71}},\ \bibinfo {pages} {933} (\bibinfo {year} {1997})}\BibitemShut {NoStop}%
\bibitem [{\citenamefont {Higashiwaki}\ \emph {et~al.}(2012)\citenamefont {Higashiwaki}, \citenamefont {Sasaki}, \citenamefont {Kuramata}, \citenamefont {Masui},\ and\ \citenamefont {Yamakoshi}}]{Shigenobu2012}%
  \BibitemOpen
  \bibfield  {author} {\bibinfo {author} {\bibfnamefont {M.}~\bibnamefont {Higashiwaki}}, \bibinfo {author} {\bibfnamefont {K.}~\bibnamefont {Sasaki}}, \bibinfo {author} {\bibfnamefont {A.}~\bibnamefont {Kuramata}}, \bibinfo {author} {\bibfnamefont {T.}~\bibnamefont {Masui}},\ and\ \bibinfo {author} {\bibfnamefont {S.}~\bibnamefont {Yamakoshi}},\ }\href {https://doi.org/10.1063/1.3674287} {\bibfield  {journal} {\bibinfo  {journal} {Applied Physics Letters}\ }\textbf {\bibinfo {volume} {100}},\ \bibinfo {pages} {013504} (\bibinfo {year} {2012})}\BibitemShut {NoStop}%
\bibitem [{\citenamefont {V{\'\i}llora}\ \emph {et~al.}(2008)\citenamefont {V{\'\i}llora}, \citenamefont {Shimamura}, \citenamefont {Ujiie},\ and\ \citenamefont {Aoki}}]{Kazuo2008}%
  \BibitemOpen
  \bibfield  {author} {\bibinfo {author} {\bibfnamefont {E.~G.}\ \bibnamefont {V{\'\i}llora}}, \bibinfo {author} {\bibfnamefont {K.}~\bibnamefont {Shimamura}}, \bibinfo {author} {\bibfnamefont {T.}~\bibnamefont {Ujiie}},\ and\ \bibinfo {author} {\bibfnamefont {K.}~\bibnamefont {Aoki}},\ }\href {https://doi.org/10.1063/1.2910770} {\bibfield  {journal} {\bibinfo  {journal} {Applied Physics Letters}\ }\textbf {\bibinfo {volume} {92}},\ \bibinfo {pages} {202118} (\bibinfo {year} {2008})}\BibitemShut {NoStop}%
\bibitem [{\citenamefont {Galazka}\ \emph {et~al.}(2014)\citenamefont {Galazka}, \citenamefont {Irmscher}, \citenamefont {Uecker}, \citenamefont {Bertram}, \citenamefont {Pietsch}, \citenamefont {Kwasniewski}, \citenamefont {Naumann}, \citenamefont {Schulz}, \citenamefont {Schewski}, \citenamefont {Klimm},\ and\ \citenamefont {Bickermann}}]{Bickermann2014}%
  \BibitemOpen
  \bibfield  {author} {\bibinfo {author} {\bibfnamefont {Z.}~\bibnamefont {Galazka}}, \bibinfo {author} {\bibfnamefont {K.}~\bibnamefont {Irmscher}}, \bibinfo {author} {\bibfnamefont {R.}~\bibnamefont {Uecker}}, \bibinfo {author} {\bibfnamefont {R.}~\bibnamefont {Bertram}}, \bibinfo {author} {\bibfnamefont {M.}~\bibnamefont {Pietsch}}, \bibinfo {author} {\bibfnamefont {A.}~\bibnamefont {Kwasniewski}}, \bibinfo {author} {\bibfnamefont {M.}~\bibnamefont {Naumann}}, \bibinfo {author} {\bibfnamefont {T.}~\bibnamefont {Schulz}}, \bibinfo {author} {\bibfnamefont {R.}~\bibnamefont {Schewski}}, \bibinfo {author} {\bibfnamefont {D.}~\bibnamefont {Klimm}},\ and\ \bibinfo {author} {\bibfnamefont {M.}~\bibnamefont {Bickermann}},\ }\href {https://doi.org/https://doi.org/10.1016/j.jcrysgro.2014.07.021} {\bibfield  {journal} {\bibinfo  {journal} {Journal of Crystal Growth}\ }\textbf {\bibinfo {volume} {404}},\ \bibinfo {pages} {184} (\bibinfo {year} {2014})}\BibitemShut {NoStop}%
\bibitem [{\citenamefont {Green}\ \emph {et~al.}(2022)\citenamefont {Green}, \citenamefont {Speck}, \citenamefont {Xing}, \citenamefont {Moens}, \citenamefont {Allerstam}, \citenamefont {Gumaelius}, \citenamefont {Neyer}, \citenamefont {Arias-Purdue}, \citenamefont {Mehrotra}, \citenamefont {Kuramata}, \citenamefont {Sasaki}, \citenamefont {Watanabe}, \citenamefont {Koshi}, \citenamefont {Blevins}, \citenamefont {Bierwagen}, \citenamefont {Krishnamoorthy}, \citenamefont {Leedy}, \citenamefont {Arehart}, \citenamefont {Neal}, \citenamefont {Mou}, \citenamefont {Ringel}, \citenamefont {Kumar}, \citenamefont {Sharma}, \citenamefont {Ghosh}, \citenamefont {Singisetti}, \citenamefont {Li}, \citenamefont {Chabak}, \citenamefont {Liddy}, \citenamefont {Islam}, \citenamefont {Rajan}, \citenamefont {Graham}, \citenamefont {Choi}, \citenamefont {Cheng},\ and\ \citenamefont {Higashiwaki}}]{GreenAIP2022}%
  \BibitemOpen
  \bibfield  {author} {\bibinfo {author} {\bibfnamefont {A.~J.}\ \bibnamefont {Green}}, \bibinfo {author} {\bibfnamefont {J.}~\bibnamefont {Speck}}, \bibinfo {author} {\bibfnamefont {G.}~\bibnamefont {Xing}}, \bibinfo {author} {\bibfnamefont {P.}~\bibnamefont {Moens}}, \bibinfo {author} {\bibfnamefont {F.}~\bibnamefont {Allerstam}}, \bibinfo {author} {\bibfnamefont {K.}~\bibnamefont {Gumaelius}}, \bibinfo {author} {\bibfnamefont {T.}~\bibnamefont {Neyer}}, \bibinfo {author} {\bibfnamefont {A.}~\bibnamefont {Arias-Purdue}}, \bibinfo {author} {\bibfnamefont {V.}~\bibnamefont {Mehrotra}}, \bibinfo {author} {\bibfnamefont {A.}~\bibnamefont {Kuramata}}, \bibinfo {author} {\bibfnamefont {K.}~\bibnamefont {Sasaki}}, \bibinfo {author} {\bibfnamefont {S.}~\bibnamefont {Watanabe}}, \bibinfo {author} {\bibfnamefont {K.}~\bibnamefont {Koshi}}, \bibinfo {author} {\bibfnamefont {J.}~\bibnamefont {Blevins}}, \bibinfo {author} {\bibfnamefont {O.}~\bibnamefont {Bierwagen}}, \bibinfo {author} {\bibfnamefont {S.}~\bibnamefont
  {Krishnamoorthy}}, \bibinfo {author} {\bibfnamefont {K.}~\bibnamefont {Leedy}}, \bibinfo {author} {\bibfnamefont {A.~R.}\ \bibnamefont {Arehart}}, \bibinfo {author} {\bibfnamefont {A.~T.}\ \bibnamefont {Neal}}, \bibinfo {author} {\bibfnamefont {S.}~\bibnamefont {Mou}}, \bibinfo {author} {\bibfnamefont {S.~A.}\ \bibnamefont {Ringel}}, \bibinfo {author} {\bibfnamefont {A.}~\bibnamefont {Kumar}}, \bibinfo {author} {\bibfnamefont {A.}~\bibnamefont {Sharma}}, \bibinfo {author} {\bibfnamefont {K.}~\bibnamefont {Ghosh}}, \bibinfo {author} {\bibfnamefont {U.}~\bibnamefont {Singisetti}}, \bibinfo {author} {\bibfnamefont {W.}~\bibnamefont {Li}}, \bibinfo {author} {\bibfnamefont {K.}~\bibnamefont {Chabak}}, \bibinfo {author} {\bibfnamefont {K.}~\bibnamefont {Liddy}}, \bibinfo {author} {\bibfnamefont {A.}~\bibnamefont {Islam}}, \bibinfo {author} {\bibfnamefont {S.}~\bibnamefont {Rajan}}, \bibinfo {author} {\bibfnamefont {S.}~\bibnamefont {Graham}}, \bibinfo {author} {\bibfnamefont {S.}~\bibnamefont {Choi}}, \bibinfo
  {author} {\bibfnamefont {Z.}~\bibnamefont {Cheng}},\ and\ \bibinfo {author} {\bibfnamefont {M.}~\bibnamefont {Higashiwaki}},\ }\href {https://doi.org/10.1063/5.0060327} {\bibfield  {journal} {\bibinfo  {journal} {APL Materials}\ }\textbf {\bibinfo {volume} {10}},\ \bibinfo {pages} {029201} (\bibinfo {year} {2022})}\BibitemShut {NoStop}%
\bibitem [{\citenamefont {Pearton}\ \emph {et~al.}(2018)\citenamefont {Pearton}, \citenamefont {Yang}, \citenamefont {Cary}, \citenamefont {Ren}, \citenamefont {Kim}, \citenamefont {Tadjer},\ and\ \citenamefont {Mastro}}]{Pearton2018}%
  \BibitemOpen
  \bibfield  {author} {\bibinfo {author} {\bibfnamefont {S.~J.}\ \bibnamefont {Pearton}}, \bibinfo {author} {\bibfnamefont {J.}~\bibnamefont {Yang}}, \bibinfo {author} {\bibfnamefont {I.}~\bibnamefont {Cary}, \bibfnamefont {Patrick~H.}}, \bibinfo {author} {\bibfnamefont {F.}~\bibnamefont {Ren}}, \bibinfo {author} {\bibfnamefont {J.}~\bibnamefont {Kim}}, \bibinfo {author} {\bibfnamefont {M.~J.}\ \bibnamefont {Tadjer}},\ and\ \bibinfo {author} {\bibfnamefont {M.~A.}\ \bibnamefont {Mastro}},\ }\href {https://doi.org/10.1063/1.5006941} {\bibfield  {journal} {\bibinfo  {journal} {Applied Physics Reviews}\ }\textbf {\bibinfo {volume} {5}},\ \bibinfo {pages} {011301} (\bibinfo {year} {2018})}\BibitemShut {NoStop}%
\bibitem [{\citenamefont {Huang}\ \emph {et~al.}(2024)\citenamefont {Huang}, \citenamefont {Wang}, \citenamefont {Zhou}, \citenamefont {Xing}, \citenamefont {Wang}, \citenamefont {Zhang}, \citenamefont {Tang}, \citenamefont {Huang},\ and\ \citenamefont {Wang}}]{HuangACS2024}%
  \BibitemOpen
  \bibfield  {author} {\bibinfo {author} {\bibfnamefont {H.}~\bibnamefont {Huang}}, \bibinfo {author} {\bibfnamefont {L.}~\bibnamefont {Wang}}, \bibinfo {author} {\bibfnamefont {H.}~\bibnamefont {Zhou}}, \bibinfo {author} {\bibfnamefont {H.}~\bibnamefont {Xing}}, \bibinfo {author} {\bibfnamefont {L.}~\bibnamefont {Wang}}, \bibinfo {author} {\bibfnamefont {W.}~\bibnamefont {Zhang}}, \bibinfo {author} {\bibfnamefont {K.}~\bibnamefont {Tang}}, \bibinfo {author} {\bibfnamefont {J.}~\bibnamefont {Huang}},\ and\ \bibinfo {author} {\bibfnamefont {L.}~\bibnamefont {Wang}},\ }\bibfield  {journal} {\bibinfo  {journal} {ACS Applied Materials \& Interfaces}\ }\href {https://doi.org/10.1021/acsami.4c15345} {10.1021/acsami.4c15345} (\bibinfo {year} {2024})\BibitemShut {NoStop}%
\bibitem [{\citenamefont {P{\'e}rez-Tom{\'a}s}\ \emph {et~al.}(2019)\citenamefont {P{\'e}rez-Tom{\'a}s}, \citenamefont {Chikoidze}, \citenamefont {Dumont}, \citenamefont {Jennings}, \citenamefont {Russell}, \citenamefont {Vales-Castro}, \citenamefont {Catalan}, \citenamefont {Lira-Cant{\'u}}, \citenamefont {{Ton --That}}, \citenamefont {Teherani}, \citenamefont {Sandana}, \citenamefont {Bove},\ and\ \citenamefont {Rogers}}]{Perez2019}%
  \BibitemOpen
  \bibfield  {author} {\bibinfo {author} {\bibfnamefont {A.}~\bibnamefont {P{\'e}rez-Tom{\'a}s}}, \bibinfo {author} {\bibfnamefont {E.}~\bibnamefont {Chikoidze}}, \bibinfo {author} {\bibfnamefont {Y.}~\bibnamefont {Dumont}}, \bibinfo {author} {\bibfnamefont {M.}~\bibnamefont {Jennings}}, \bibinfo {author} {\bibfnamefont {S.}~\bibnamefont {Russell}}, \bibinfo {author} {\bibfnamefont {P.}~\bibnamefont {Vales-Castro}}, \bibinfo {author} {\bibfnamefont {G.}~\bibnamefont {Catalan}}, \bibinfo {author} {\bibfnamefont {M.}~\bibnamefont {Lira-Cant{\'u}}}, \bibinfo {author} {\bibfnamefont {C.}~\bibnamefont {{Ton --That}}}, \bibinfo {author} {\bibfnamefont {F.}~\bibnamefont {Teherani}}, \bibinfo {author} {\bibfnamefont {V.}~\bibnamefont {Sandana}}, \bibinfo {author} {\bibfnamefont {P.}~\bibnamefont {Bove}},\ and\ \bibinfo {author} {\bibfnamefont {D.}~\bibnamefont {Rogers}},\ }\href {https://doi.org/https://doi.org/10.1016/j.mtener.2019.100350} {\bibfield  {journal} {\bibinfo  {journal} {Materials Today Energy}\ }\textbf
  {\bibinfo {volume} {14}},\ \bibinfo {pages} {100350} (\bibinfo {year} {2019})}\BibitemShut {NoStop}%
\bibitem [{\citenamefont {Giri}\ \emph {et~al.}(2024)\citenamefont {Giri}, \citenamefont {Mahata}, \citenamefont {Guha},\ and\ \citenamefont {Banerji}}]{Giri2024}%
  \BibitemOpen
  \bibfield  {author} {\bibinfo {author} {\bibfnamefont {S.}~\bibnamefont {Giri}}, \bibinfo {author} {\bibfnamefont {B.}~\bibnamefont {Mahata}}, \bibinfo {author} {\bibfnamefont {P.~K.}\ \bibnamefont {Guha}},\ and\ \bibinfo {author} {\bibfnamefont {P.}~\bibnamefont {Banerji}},\ }\href {https://doi.org/10.1021/acsaelm.3c01187} {\bibfield  {journal} {\bibinfo  {journal} {ACS Applied Electronic Materials}\ }\textbf {\bibinfo {volume} {6}},\ \bibinfo {pages} {230} (\bibinfo {year} {2024})}\BibitemShut {NoStop}%
\bibitem [{\citenamefont {Aida}\ \emph {et~al.}(2008)\citenamefont {Aida}, \citenamefont {Nishiguchi}, \citenamefont {Takeda}, \citenamefont {Aota}, \citenamefont {Sunakawa},\ and\ \citenamefont {Yaguchi}}]{Aida_2008}%
  \BibitemOpen
  \bibfield  {author} {\bibinfo {author} {\bibfnamefont {H.}~\bibnamefont {Aida}}, \bibinfo {author} {\bibfnamefont {K.}~\bibnamefont {Nishiguchi}}, \bibinfo {author} {\bibfnamefont {H.}~\bibnamefont {Takeda}}, \bibinfo {author} {\bibfnamefont {N.}~\bibnamefont {Aota}}, \bibinfo {author} {\bibfnamefont {K.}~\bibnamefont {Sunakawa}},\ and\ \bibinfo {author} {\bibfnamefont {Y.}~\bibnamefont {Yaguchi}},\ }\href {https://doi.org/10.1143/JJAP.47.8506} {\bibfield  {journal} {\bibinfo  {journal} {Japanese Journal of Applied Physics}\ }\textbf {\bibinfo {volume} {47}},\ \bibinfo {pages} {8506} (\bibinfo {year} {2008})}\BibitemShut {NoStop}%
\bibitem [{\citenamefont {Feng}\ \emph {et~al.}(2024)\citenamefont {Feng}, \citenamefont {Li}, \citenamefont {Tian}, \citenamefont {Qi}, \citenamefont {Guo},\ and\ \citenamefont {Tang}}]{Weihua2024}%
  \BibitemOpen
  \bibfield  {author} {\bibinfo {author} {\bibfnamefont {G.}~\bibnamefont {Feng}}, \bibinfo {author} {\bibfnamefont {S.}~\bibnamefont {Li}}, \bibinfo {author} {\bibfnamefont {Y.}~\bibnamefont {Tian}}, \bibinfo {author} {\bibfnamefont {S.}~\bibnamefont {Qi}}, \bibinfo {author} {\bibfnamefont {D.}~\bibnamefont {Guo}},\ and\ \bibinfo {author} {\bibfnamefont {W.}~\bibnamefont {Tang}},\ }\href {https://doi.org/10.1021/acsomega.4c00405} {\bibfield  {journal} {\bibinfo  {journal} {ACS Omega}\ }\textbf {\bibinfo {volume} {9}},\ \bibinfo {pages} {22084} (\bibinfo {year} {2024})}\BibitemShut {NoStop}%
\bibitem [{\citenamefont {Takeuchi}\ \emph {et~al.}(2008)\citenamefont {Takeuchi}, \citenamefont {Ishikawa}, \citenamefont {Takeuchi},\ and\ \citenamefont {Horikoshi}}]{Takeuchi2008}%
  \BibitemOpen
  \bibfield  {author} {\bibinfo {author} {\bibfnamefont {T.}~\bibnamefont {Takeuchi}}, \bibinfo {author} {\bibfnamefont {H.}~\bibnamefont {Ishikawa}}, \bibinfo {author} {\bibfnamefont {N.}~\bibnamefont {Takeuchi}},\ and\ \bibinfo {author} {\bibfnamefont {Y.}~\bibnamefont {Horikoshi}},\ }\href {https://doi.org/https://doi.org/10.1016/j.tsf.2007.06.075} {\bibfield  {journal} {\bibinfo  {journal} {Thin Solid Films}\ }\textbf {\bibinfo {volume} {516}},\ \bibinfo {pages} {4593} (\bibinfo {year} {2008})},\ \bibinfo {note} {6th International Conference on Coatings on Glass and Plastics (ICCG6)- Advanced Coatings for Large-Area or High-Volume Products-}\BibitemShut {NoStop}%
\bibitem [{\citenamefont {Higashiwaki}\ \emph {et~al.}(2014)\citenamefont {Higashiwaki}, \citenamefont {Sasaki}, \citenamefont {Kuramata}, \citenamefont {Masui},\ and\ \citenamefont {Yamakoshi}}]{Higashiwaki2014}%
  \BibitemOpen
  \bibfield  {author} {\bibinfo {author} {\bibfnamefont {M.}~\bibnamefont {Higashiwaki}}, \bibinfo {author} {\bibfnamefont {K.}~\bibnamefont {Sasaki}}, \bibinfo {author} {\bibfnamefont {A.}~\bibnamefont {Kuramata}}, \bibinfo {author} {\bibfnamefont {T.}~\bibnamefont {Masui}},\ and\ \bibinfo {author} {\bibfnamefont {S.}~\bibnamefont {Yamakoshi}},\ }\href {https://doi.org/https://doi.org/10.1002/pssa.201330197} {\bibfield  {journal} {\bibinfo  {journal} {physica status solidi (a)}\ }\textbf {\bibinfo {volume} {211}},\ \bibinfo {pages} {21} (\bibinfo {year} {2014})}\BibitemShut {NoStop}%
\bibitem [{\citenamefont {Orita}\ \emph {et~al.}(2000)\citenamefont {Orita}, \citenamefont {Ohta}, \citenamefont {Hirano},\ and\ \citenamefont {Hosono}}]{Orita2000}%
  \BibitemOpen
  \bibfield  {author} {\bibinfo {author} {\bibfnamefont {M.}~\bibnamefont {Orita}}, \bibinfo {author} {\bibfnamefont {H.}~\bibnamefont {Ohta}}, \bibinfo {author} {\bibfnamefont {M.}~\bibnamefont {Hirano}},\ and\ \bibinfo {author} {\bibfnamefont {H.}~\bibnamefont {Hosono}},\ }\href {https://doi.org/10.1063/1.1330559} {\bibfield  {journal} {\bibinfo  {journal} {Applied Physics Letters}\ }\textbf {\bibinfo {volume} {77}},\ \bibinfo {pages} {4166} (\bibinfo {year} {2000})}\BibitemShut {NoStop}%
\bibitem [{\citenamefont {Motti}\ \emph {et~al.}(2021)\citenamefont {Motti}, \citenamefont {Patel}, \citenamefont {Oliver}, \citenamefont {Snaith}, \citenamefont {Johnston},\ and\ \citenamefont {Herz}}]{Motti2021}%
  \BibitemOpen
  \bibfield  {author} {\bibinfo {author} {\bibfnamefont {S.~G.}\ \bibnamefont {Motti}}, \bibinfo {author} {\bibfnamefont {J.~B.}\ \bibnamefont {Patel}}, \bibinfo {author} {\bibfnamefont {R.~D.~J.}\ \bibnamefont {Oliver}}, \bibinfo {author} {\bibfnamefont {H.~J.}\ \bibnamefont {Snaith}}, \bibinfo {author} {\bibfnamefont {M.~B.}\ \bibnamefont {Johnston}},\ and\ \bibinfo {author} {\bibfnamefont {L.~M.}\ \bibnamefont {Herz}},\ }\href {https://doi.org/10.1038/s41467-021-26930-4} {\bibfield  {journal} {\bibinfo  {journal} {Nature Communications}\ }\textbf {\bibinfo {volume} {12}},\ \bibinfo {pages} {6955} (\bibinfo {year} {2021})}\BibitemShut {NoStop}%
\bibitem [{\citenamefont {Kabir}\ and\ \citenamefont {Demirocak}(2017)}]{Kabir2017}%
  \BibitemOpen
  \bibfield  {author} {\bibinfo {author} {\bibfnamefont {M.~M.}\ \bibnamefont {Kabir}}\ and\ \bibinfo {author} {\bibfnamefont {D.~E.}\ \bibnamefont {Demirocak}},\ }\href {https://doi.org/https://doi.org/10.1002/er.3762} {\bibfield  {journal} {\bibinfo  {journal} {International Journal of Energy Research}\ }\textbf {\bibinfo {volume} {41}},\ \bibinfo {pages} {1963} (\bibinfo {year} {2017})}\BibitemShut {NoStop}%
\bibitem [{\citenamefont {Jeong}\ \emph {et~al.}(2013)\citenamefont {Jeong}, \citenamefont {Aetukuri}, \citenamefont {Graf}, \citenamefont {Schladt}, \citenamefont {Samant},\ and\ \citenamefont {Parkin}}]{Jeong2013}%
  \BibitemOpen
  \bibfield  {author} {\bibinfo {author} {\bibfnamefont {J.}~\bibnamefont {Jeong}}, \bibinfo {author} {\bibfnamefont {N.}~\bibnamefont {Aetukuri}}, \bibinfo {author} {\bibfnamefont {T.}~\bibnamefont {Graf}}, \bibinfo {author} {\bibfnamefont {T.~D.}\ \bibnamefont {Schladt}}, \bibinfo {author} {\bibfnamefont {M.~G.}\ \bibnamefont {Samant}},\ and\ \bibinfo {author} {\bibfnamefont {S.~S.~P.}\ \bibnamefont {Parkin}},\ }\href {https://doi.org/10.1126/science.1230512} {\bibfield  {journal} {\bibinfo  {journal} {Science}\ }\textbf {\bibinfo {volume} {339}},\ \bibinfo {pages} {1402} (\bibinfo {year} {2013})}\BibitemShut {NoStop}%
\bibitem [{\citenamefont {Meng}\ \emph {et~al.}(2024)\citenamefont {Meng}, \citenamefont {Sheikh}, \citenamefont {Jacobs}, \citenamefont {Liu}, \citenamefont {Nachlas}, \citenamefont {Li},\ and\ \citenamefont {Morgan}}]{Meng2024}%
  \BibitemOpen
  \bibfield  {author} {\bibinfo {author} {\bibfnamefont {J.}~\bibnamefont {Meng}}, \bibinfo {author} {\bibfnamefont {M.~S.}\ \bibnamefont {Sheikh}}, \bibinfo {author} {\bibfnamefont {R.}~\bibnamefont {Jacobs}}, \bibinfo {author} {\bibfnamefont {J.}~\bibnamefont {Liu}}, \bibinfo {author} {\bibfnamefont {W.~O.}\ \bibnamefont {Nachlas}}, \bibinfo {author} {\bibfnamefont {X.}~\bibnamefont {Li}},\ and\ \bibinfo {author} {\bibfnamefont {D.}~\bibnamefont {Morgan}},\ }\href {https://doi.org/10.1038/s41563-024-01919-8} {\bibfield  {journal} {\bibinfo  {journal} {Nature Materials}\ }\textbf {\bibinfo {volume} {23}},\ \bibinfo {pages} {1252} (\bibinfo {year} {2024})}\BibitemShut {NoStop}%
\bibitem [{\citenamefont {Zimmermann}\ \emph {et~al.}(2020)\citenamefont {Zimmermann}, \citenamefont {R\o{}nning}, \citenamefont {Kalmann~Frodason}, \citenamefont {Bobal}, \citenamefont {Vines},\ and\ \citenamefont {Varley}}]{18_Zimmermann_2020}%
  \BibitemOpen
  \bibfield  {author} {\bibinfo {author} {\bibfnamefont {C.}~\bibnamefont {Zimmermann}}, \bibinfo {author} {\bibfnamefont {V.}~\bibnamefont {R\o{}nning}}, \bibinfo {author} {\bibfnamefont {Y.}~\bibnamefont {Kalmann~Frodason}}, \bibinfo {author} {\bibfnamefont {V.}~\bibnamefont {Bobal}}, \bibinfo {author} {\bibfnamefont {L.}~\bibnamefont {Vines}},\ and\ \bibinfo {author} {\bibfnamefont {J.~B.}\ \bibnamefont {Varley}},\ }\href {https://doi.org/10.1103/PhysRevMaterials.4.074605} {\bibfield  {journal} {\bibinfo  {journal} {Phys. Rev. Mater.}\ }\textbf {\bibinfo {volume} {4}},\ \bibinfo {pages} {074605} (\bibinfo {year} {2020})}\BibitemShut {NoStop}%
\bibitem [{\citenamefont {Jeong}\ \emph {et~al.}(2022)\citenamefont {Jeong}, \citenamefont {Ertekin},\ and\ \citenamefont {Seebauer}}]{JeongACS2022}%
  \BibitemOpen
  \bibfield  {author} {\bibinfo {author} {\bibfnamefont {H.}~\bibnamefont {Jeong}}, \bibinfo {author} {\bibfnamefont {E.}~\bibnamefont {Ertekin}},\ and\ \bibinfo {author} {\bibfnamefont {E.~G.}\ \bibnamefont {Seebauer}},\ }\href {https://doi.org/10.1021/acsami.2c07672} {\bibfield  {journal} {\bibinfo  {journal} {ACS Applied Materials \& Interfaces}\ }\textbf {\bibinfo {volume} {14}},\ \bibinfo {pages} {34059} (\bibinfo {year} {2022})}\BibitemShut {NoStop}%
\bibitem [{\citenamefont {Uhlendorf}\ \emph {et~al.}(2021)\citenamefont {Uhlendorf}, \citenamefont {Galazka},\ and\ \citenamefont {Schmidt}}]{Uhlendorf2021}%
  \BibitemOpen
  \bibfield  {author} {\bibinfo {author} {\bibfnamefont {J.}~\bibnamefont {Uhlendorf}}, \bibinfo {author} {\bibfnamefont {Z.}~\bibnamefont {Galazka}},\ and\ \bibinfo {author} {\bibfnamefont {H.}~\bibnamefont {Schmidt}},\ }\href {https://doi.org/10.1063/5.0071729} {\bibfield  {journal} {\bibinfo  {journal} {Applied Physics Letters}\ }\textbf {\bibinfo {volume} {119}},\ \bibinfo {pages} {242106} (\bibinfo {year} {2021})}\BibitemShut {NoStop}%
\bibitem [{\citenamefont {Uhlendorf}\ and\ \citenamefont {Schmidt}(2023)}]{Uhlendorf2023}%
  \BibitemOpen
  \bibfield  {author} {\bibinfo {author} {\bibfnamefont {J.}~\bibnamefont {Uhlendorf}}\ and\ \bibinfo {author} {\bibfnamefont {H.}~\bibnamefont {Schmidt}},\ }\href {https://doi.org/10.1103/PhysRevMaterials.7.093402} {\bibfield  {journal} {\bibinfo  {journal} {Phys. Rev. Mater.}\ }\textbf {\bibinfo {volume} {7}},\ \bibinfo {pages} {093402} (\bibinfo {year} {2023})}\BibitemShut {NoStop}%
\bibitem [{\citenamefont {Uhlendorf}\ and\ \citenamefont {Schmidt}(2024)}]{Uhlendorf2024}%
  \BibitemOpen
  \bibfield  {author} {\bibinfo {author} {\bibfnamefont {J.}~\bibnamefont {Uhlendorf}}\ and\ \bibinfo {author} {\bibfnamefont {H.}~\bibnamefont {Schmidt}},\ }\href {https://doi.org/doi:10.1515/znb-2023-0091} {\bibfield  {journal} {\bibinfo  {journal} {Zeitschrift f{\"u}r Naturforschung B}\ }\textbf {\bibinfo {volume} {79}},\ \bibinfo {pages} {225} (\bibinfo {year} {2024})}\BibitemShut {NoStop}%
\bibitem [{\citenamefont {Kohan}\ \emph {et~al.}(2000)\citenamefont {Kohan}, \citenamefont {Ceder}, \citenamefont {Morgan},\ and\ \citenamefont {Van~de Walle}}]{Kohan2000}%
  \BibitemOpen
  \bibfield  {author} {\bibinfo {author} {\bibfnamefont {A.~F.}\ \bibnamefont {Kohan}}, \bibinfo {author} {\bibfnamefont {G.}~\bibnamefont {Ceder}}, \bibinfo {author} {\bibfnamefont {D.}~\bibnamefont {Morgan}},\ and\ \bibinfo {author} {\bibfnamefont {C.~G.}\ \bibnamefont {Van~de Walle}},\ }\href {https://doi.org/10.1103/PhysRevB.61.15019} {\bibfield  {journal} {\bibinfo  {journal} {Phys. Rev. B}\ }\textbf {\bibinfo {volume} {61}},\ \bibinfo {pages} {15019} (\bibinfo {year} {2000})}\BibitemShut {NoStop}%
\bibitem [{\citenamefont {Peelaers}\ \emph {et~al.}(2019)\citenamefont {Peelaers}, \citenamefont {Lyons}, \citenamefont {Varley},\ and\ \citenamefont {Van~de Walle}}]{Peelaers2019}%
  \BibitemOpen
  \bibfield  {author} {\bibinfo {author} {\bibfnamefont {H.}~\bibnamefont {Peelaers}}, \bibinfo {author} {\bibfnamefont {J.~L.}\ \bibnamefont {Lyons}}, \bibinfo {author} {\bibfnamefont {J.~B.}\ \bibnamefont {Varley}},\ and\ \bibinfo {author} {\bibfnamefont {C.~G.}\ \bibnamefont {Van~de Walle}},\ }\bibfield  {journal} {\bibinfo  {journal} {APL Materials}\ }\textbf {\bibinfo {volume} {7}},\ \href {https://doi.org/10.1063/1.5063807} {10.1063/1.5063807} (\bibinfo {year} {2019}),\ \bibinfo {note} {022519}\BibitemShut {NoStop}%
\bibitem [{\citenamefont {Lee}\ \emph {et~al.}(2023)\citenamefont {Lee}, \citenamefont {Rock}, \citenamefont {Islam}, \citenamefont {Scarpulla},\ and\ \citenamefont {Ertekin}}]{0_Lee_2023}%
  \BibitemOpen
  \bibfield  {author} {\bibinfo {author} {\bibfnamefont {C.}~\bibnamefont {Lee}}, \bibinfo {author} {\bibfnamefont {N.~D.}\ \bibnamefont {Rock}}, \bibinfo {author} {\bibfnamefont {A.}~\bibnamefont {Islam}}, \bibinfo {author} {\bibfnamefont {M.~A.}\ \bibnamefont {Scarpulla}},\ and\ \bibinfo {author} {\bibfnamefont {E.}~\bibnamefont {Ertekin}},\ }\bibfield  {journal} {\bibinfo  {journal} {APL Materials}\ }\textbf {\bibinfo {volume} {11}},\ \href {https://doi.org/10.1063/5.0131453} {10.1063/5.0131453} (\bibinfo {year} {2023}),\ \bibinfo {note} {011106}\BibitemShut {NoStop}%
\bibitem [{\citenamefont {Alberi}\ and\ \citenamefont {Scarpulla}(2016)}]{Alberi2016}%
  \BibitemOpen
  \bibfield  {author} {\bibinfo {author} {\bibfnamefont {K.}~\bibnamefont {Alberi}}\ and\ \bibinfo {author} {\bibfnamefont {M.}~\bibnamefont {Scarpulla}},\ }\bibfield  {journal} {\bibinfo  {journal} {Scientific Reports}\ }\textbf {\bibinfo {volume} {6}},\ \href {https://doi.org/10.1038/srep27954} {10.1038/srep27954} (\bibinfo {year} {2016})\BibitemShut {NoStop}%
\bibitem [{\citenamefont {Zhang}\ \emph {et~al.}(2001)\citenamefont {Zhang}, \citenamefont {Wei},\ and\ \citenamefont {Zunger}}]{Zunger2001}%
  \BibitemOpen
  \bibfield  {author} {\bibinfo {author} {\bibfnamefont {S.~B.}\ \bibnamefont {Zhang}}, \bibinfo {author} {\bibfnamefont {S.-H.}\ \bibnamefont {Wei}},\ and\ \bibinfo {author} {\bibfnamefont {A.}~\bibnamefont {Zunger}},\ }\href {https://doi.org/10.1103/PhysRevB.63.075205} {\bibfield  {journal} {\bibinfo  {journal} {Phys. Rev. B}\ }\textbf {\bibinfo {volume} {63}},\ \bibinfo {pages} {075205} (\bibinfo {year} {2001})}\BibitemShut {NoStop}%
\bibitem [{\citenamefont {Moses}\ \emph {et~al.}(2016)\citenamefont {Moses}, \citenamefont {Janotti}, \citenamefont {Franchini}, \citenamefont {Kresse},\ and\ \citenamefont {Van~de Walle}}]{Moses2016}%
  \BibitemOpen
  \bibfield  {author} {\bibinfo {author} {\bibfnamefont {P.~G.}\ \bibnamefont {Moses}}, \bibinfo {author} {\bibfnamefont {A.}~\bibnamefont {Janotti}}, \bibinfo {author} {\bibfnamefont {C.}~\bibnamefont {Franchini}}, \bibinfo {author} {\bibfnamefont {G.}~\bibnamefont {Kresse}},\ and\ \bibinfo {author} {\bibfnamefont {C.~G.}\ \bibnamefont {Van~de Walle}},\ }\href {https://doi.org/10.1063/1.4948239} {\bibfield  {journal} {\bibinfo  {journal} {Journal of Applied Physics}\ }\textbf {\bibinfo {volume} {119}},\ \bibinfo {pages} {181503} (\bibinfo {year} {2016})}\BibitemShut {NoStop}%
\bibitem [{\citenamefont {Jeong}\ and\ \citenamefont {Seebauer}(2023)}]{JeongJVSTA2023}%
  \BibitemOpen
  \bibfield  {author} {\bibinfo {author} {\bibfnamefont {H.}~\bibnamefont {Jeong}}\ and\ \bibinfo {author} {\bibfnamefont {E.~G.}\ \bibnamefont {Seebauer}},\ }\href {https://doi.org/10.1116/6.0002467} {\bibfield  {journal} {\bibinfo  {journal} {Journal of Vacuum Science \& Technology A}\ }\textbf {\bibinfo {volume} {41}},\ \bibinfo {pages} {033203} (\bibinfo {year} {2023})}\BibitemShut {NoStop}%
\bibitem [{\citenamefont {Lee}\ \emph {et~al.}(2024)\citenamefont {Lee}, \citenamefont {Scarpulla},\ and\ \citenamefont {Ertekin}}]{CLee2024}%
  \BibitemOpen
  \bibfield  {author} {\bibinfo {author} {\bibfnamefont {C.}~\bibnamefont {Lee}}, \bibinfo {author} {\bibfnamefont {M.~A.}\ \bibnamefont {Scarpulla}},\ and\ \bibinfo {author} {\bibfnamefont {E.}~\bibnamefont {Ertekin}},\ }\href {https://doi.org/10.1103/PhysRevMaterials.8.054603} {\bibfield  {journal} {\bibinfo  {journal} {Phys. Rev. Mater.}\ }\textbf {\bibinfo {volume} {8}},\ \bibinfo {pages} {054603} (\bibinfo {year} {2024})}\BibitemShut {NoStop}%
\bibitem [{\citenamefont {Trinkle}(2016)}]{20_Trinkle_2016}%
  \BibitemOpen
  \bibfield  {author} {\bibinfo {author} {\bibfnamefont {D.~R.}\ \bibnamefont {Trinkle}},\ }\href {https://doi.org/10.1080/14786435.2016.1212175} {\bibfield  {journal} {\bibinfo  {journal} {Philosophical Magazine}\ }\textbf {\bibinfo {volume} {96}},\ \bibinfo {pages} {2714} (\bibinfo {year} {2016})}\BibitemShut {NoStop}%
\bibitem [{\citenamefont {Trinkle}(2017)}]{21_Trinkle_2017}%
  \BibitemOpen
  \bibfield  {author} {\bibinfo {author} {\bibfnamefont {D.~R.}\ \bibnamefont {Trinkle}},\ }\href {https://doi.org/10.1080/14786435.2017.1340685} {\bibfield  {journal} {\bibinfo  {journal} {Philosophical Magazine}\ }\textbf {\bibinfo {volume} {97}},\ \bibinfo {pages} {2514} (\bibinfo {year} {2017})}\BibitemShut {NoStop}%
\bibitem [{\citenamefont {Trinkle}\ and\ \citenamefont {Jain}(2019)}]{22_Trinkle_Onsager}%
  \BibitemOpen
  \bibfield  {author} {\bibinfo {author} {\bibfnamefont {D.}~\bibnamefont {Trinkle}}\ and\ \bibinfo {author} {\bibfnamefont {A.}~\bibnamefont {Jain}},\ }\href {https://doi.org/10.5281/zenodo.3355730} {\bibinfo {title} {Dallastrinkle/onsager: Onsager v1.3.3}} (\bibinfo {year} {2019})\BibitemShut {NoStop}%
\bibitem [{\citenamefont {Honrao}\ \emph {et~al.}(2020)\citenamefont {Honrao}, \citenamefont {Rizzardi}, \citenamefont {Maa\ss{}}, \citenamefont {Trinkle},\ and\ \citenamefont {Hennig}}]{22.5_Trinkle_Onsager}%
  \BibitemOpen
  \bibfield  {author} {\bibinfo {author} {\bibfnamefont {S.~J.}\ \bibnamefont {Honrao}}, \bibinfo {author} {\bibfnamefont {Q.}~\bibnamefont {Rizzardi}}, \bibinfo {author} {\bibfnamefont {R.}~\bibnamefont {Maa\ss{}}}, \bibinfo {author} {\bibfnamefont {D.~R.}\ \bibnamefont {Trinkle}},\ and\ \bibinfo {author} {\bibfnamefont {R.~G.}\ \bibnamefont {Hennig}},\ }\href {https://doi.org/10.1103/PhysRevMaterials.4.103608} {\bibfield  {journal} {\bibinfo  {journal} {Phys. Rev. Mater.}\ }\textbf {\bibinfo {volume} {4}},\ \bibinfo {pages} {103608} (\bibinfo {year} {2020})}\BibitemShut {NoStop}%
\bibitem [{\citenamefont {Goyal}\ \emph {et~al.}(2017)\citenamefont {Goyal}, \citenamefont {Gorai}, \citenamefont {Peng}, \citenamefont {Lany},\ and\ \citenamefont {Stevanović}}]{36_Goyal_2017}%
  \BibitemOpen
  \bibfield  {author} {\bibinfo {author} {\bibfnamefont {A.}~\bibnamefont {Goyal}}, \bibinfo {author} {\bibfnamefont {P.}~\bibnamefont {Gorai}}, \bibinfo {author} {\bibfnamefont {H.}~\bibnamefont {Peng}}, \bibinfo {author} {\bibfnamefont {S.}~\bibnamefont {Lany}},\ and\ \bibinfo {author} {\bibfnamefont {V.}~\bibnamefont {Stevanović}},\ }\href {https://doi.org/https://doi.org/10.1016/j.commatsci.2016.12.040} {\bibfield  {journal} {\bibinfo  {journal} {Computational Materials Science}\ }\textbf {\bibinfo {volume} {130}},\ \bibinfo {pages} {1} (\bibinfo {year} {2017})}\BibitemShut {NoStop}%
\bibitem [{\citenamefont {Ingebrigtsen}\ \emph {et~al.}(2018)\citenamefont {Ingebrigtsen}, \citenamefont {Kuznetsov}, \citenamefont {Svensson}, \citenamefont {Alfieri}, \citenamefont {Mihaila}, \citenamefont {Badstübner}, \citenamefont {Perron}, \citenamefont {Vines},\ and\ \citenamefont {Varley}}]{17_Ingebrigtsen_2018}%
  \BibitemOpen
  \bibfield  {author} {\bibinfo {author} {\bibfnamefont {M.~E.}\ \bibnamefont {Ingebrigtsen}}, \bibinfo {author} {\bibfnamefont {A.~Y.}\ \bibnamefont {Kuznetsov}}, \bibinfo {author} {\bibfnamefont {B.~G.}\ \bibnamefont {Svensson}}, \bibinfo {author} {\bibfnamefont {G.}~\bibnamefont {Alfieri}}, \bibinfo {author} {\bibfnamefont {A.}~\bibnamefont {Mihaila}}, \bibinfo {author} {\bibfnamefont {U.}~\bibnamefont {Badstübner}}, \bibinfo {author} {\bibfnamefont {A.}~\bibnamefont {Perron}}, \bibinfo {author} {\bibfnamefont {L.}~\bibnamefont {Vines}},\ and\ \bibinfo {author} {\bibfnamefont {J.~B.}\ \bibnamefont {Varley}},\ }\bibfield  {journal} {\bibinfo  {journal} {APL Materials}\ }\textbf {\bibinfo {volume} {7}},\ \href {https://doi.org/10.1063/1.5054826} {10.1063/1.5054826} (\bibinfo {year} {2018}),\ \bibinfo {note} {022510}\BibitemShut {NoStop}%
\bibitem [{\citenamefont {{Uddin Jewel}}\ \emph {et~al.}(2023)\citenamefont {{Uddin Jewel}}, \citenamefont {Hasan},\ and\ \citenamefont {Ahmad}}]{Jewel2023}%
  \BibitemOpen
  \bibfield  {author} {\bibinfo {author} {\bibfnamefont {M.}~\bibnamefont {{Uddin Jewel}}}, \bibinfo {author} {\bibfnamefont {S.}~\bibnamefont {Hasan}},\ and\ \bibinfo {author} {\bibfnamefont {I.}~\bibnamefont {Ahmad}},\ }\href {https://doi.org/https://doi.org/10.1016/j.commatsci.2022.111950} {\bibfield  {journal} {\bibinfo  {journal} {Computational Materials Science}\ }\textbf {\bibinfo {volume} {218}},\ \bibinfo {pages} {111950} (\bibinfo {year} {2023})}\BibitemShut {NoStop}%
\bibitem [{\citenamefont {Huang}\ \emph {et~al.}(2023)\citenamefont {Huang}, \citenamefont {Xu}, \citenamefont {Yang}, \citenamefont {Yu}, \citenamefont {Wei}, \citenamefont {Ying}, \citenamefont {Liu}, \citenamefont {Jing}, \citenamefont {Li},\ and\ \citenamefont {Li}}]{Huang2023}%
  \BibitemOpen
  \bibfield  {author} {\bibinfo {author} {\bibfnamefont {Y.}~\bibnamefont {Huang}}, \bibinfo {author} {\bibfnamefont {X.}~\bibnamefont {Xu}}, \bibinfo {author} {\bibfnamefont {J.}~\bibnamefont {Yang}}, \bibinfo {author} {\bibfnamefont {X.}~\bibnamefont {Yu}}, \bibinfo {author} {\bibfnamefont {Y.}~\bibnamefont {Wei}}, \bibinfo {author} {\bibfnamefont {T.}~\bibnamefont {Ying}}, \bibinfo {author} {\bibfnamefont {Z.}~\bibnamefont {Liu}}, \bibinfo {author} {\bibfnamefont {Y.}~\bibnamefont {Jing}}, \bibinfo {author} {\bibfnamefont {W.}~\bibnamefont {Li}},\ and\ \bibinfo {author} {\bibfnamefont {X.}~\bibnamefont {Li}},\ }\href {https://doi.org/https://doi.org/10.1016/j.mtcomm.2023.105898} {\bibfield  {journal} {\bibinfo  {journal} {Materials Today Communications}\ }\textbf {\bibinfo {volume} {35}},\ \bibinfo {pages} {105898} (\bibinfo {year} {2023})}\BibitemShut {NoStop}%
\bibitem [{\citenamefont {Hohenberg}\ and\ \citenamefont {Kohn}(1964)}]{Hohenberg1964}%
  \BibitemOpen
  \bibfield  {author} {\bibinfo {author} {\bibfnamefont {P.}~\bibnamefont {Hohenberg}}\ and\ \bibinfo {author} {\bibfnamefont {W.}~\bibnamefont {Kohn}},\ }\href {https://doi.org/10.1103/PhysRev.136.B864} {\bibfield  {journal} {\bibinfo  {journal} {Physical Review}\ }\textbf {\bibinfo {volume} {136}},\ \bibinfo {pages} {B864} (\bibinfo {year} {1964})}\BibitemShut {NoStop}%
\bibitem [{\citenamefont {Kohn}\ and\ \citenamefont {Sham}(1965)}]{Kohn1965}%
  \BibitemOpen
  \bibfield  {author} {\bibinfo {author} {\bibfnamefont {W.}~\bibnamefont {Kohn}}\ and\ \bibinfo {author} {\bibfnamefont {L.~J.}\ \bibnamefont {Sham}},\ }\href {https://doi.org/10.1103/PhysRev.140.A1133} {\bibfield  {journal} {\bibinfo  {journal} {Physical Review}\ }\textbf {\bibinfo {volume} {140}},\ \bibinfo {pages} {A1133} (\bibinfo {year} {1965})}\BibitemShut {NoStop}%
\bibitem [{\citenamefont {Bl\"ochl}(1994)}]{25_Bloch_1994}%
  \BibitemOpen
  \bibfield  {author} {\bibinfo {author} {\bibfnamefont {P.~E.}\ \bibnamefont {Bl\"ochl}},\ }\href {https://doi.org/10.1103/PhysRevB.50.17953} {\bibfield  {journal} {\bibinfo  {journal} {Phys. Rev. B}\ }\textbf {\bibinfo {volume} {50}},\ \bibinfo {pages} {17953} (\bibinfo {year} {1994})}\BibitemShut {NoStop}%
\bibitem [{\citenamefont {Kresse}\ and\ \citenamefont {Joubert}(1999)}]{26_Kresse_1999}%
  \BibitemOpen
  \bibfield  {author} {\bibinfo {author} {\bibfnamefont {G.}~\bibnamefont {Kresse}}\ and\ \bibinfo {author} {\bibfnamefont {D.}~\bibnamefont {Joubert}},\ }\href {https://doi.org/10.1103/PhysRevB.59.1758} {\bibfield  {journal} {\bibinfo  {journal} {Phys. Rev. B}\ }\textbf {\bibinfo {volume} {59}},\ \bibinfo {pages} {1758} (\bibinfo {year} {1999})}\BibitemShut {NoStop}%
\bibitem [{\citenamefont {Kresse}\ and\ \citenamefont {Furthm{\"{u}}ller}(1996{\natexlab{a}})}]{Kresse1996a}%
  \BibitemOpen
  \bibfield  {author} {\bibinfo {author} {\bibfnamefont {G.}~\bibnamefont {Kresse}}\ and\ \bibinfo {author} {\bibfnamefont {J.}~\bibnamefont {Furthm{\"{u}}ller}},\ }\href {https://doi.org/10.1103/PhysRevB.54.11169} {\bibfield  {journal} {\bibinfo  {journal} {Physical Review B}\ }\textbf {\bibinfo {volume} {54}},\ \bibinfo {pages} {11169} (\bibinfo {year} {1996}{\natexlab{a}})}\BibitemShut {NoStop}%
\bibitem [{\citenamefont {Kresse}\ and\ \citenamefont {Furthm{\"{u}}ller}(1996{\natexlab{b}})}]{Kresse1996}%
  \BibitemOpen
  \bibfield  {author} {\bibinfo {author} {\bibfnamefont {G.}~\bibnamefont {Kresse}}\ and\ \bibinfo {author} {\bibfnamefont {J.}~\bibnamefont {Furthm{\"{u}}ller}},\ }\href {https://doi.org/10.1016/0927-0256(96)00008-0} {\bibfield  {journal} {\bibinfo  {journal} {Computational Materials Science}\ }\textbf {\bibinfo {volume} {6}},\ \bibinfo {pages} {15} (\bibinfo {year} {1996}{\natexlab{b}})}\BibitemShut {NoStop}%
\bibitem [{\citenamefont {Perdew}\ \emph {et~al.}(1996)\citenamefont {Perdew}, \citenamefont {Burke},\ and\ \citenamefont {Ernzerhof}}]{29_PBE_1996}%
  \BibitemOpen
  \bibfield  {author} {\bibinfo {author} {\bibfnamefont {J.~P.}\ \bibnamefont {Perdew}}, \bibinfo {author} {\bibfnamefont {K.}~\bibnamefont {Burke}},\ and\ \bibinfo {author} {\bibfnamefont {M.}~\bibnamefont {Ernzerhof}},\ }\href {https://doi.org/10.1103/PhysRevLett.77.3865} {\bibfield  {journal} {\bibinfo  {journal} {Phys. Rev. Lett.}\ }\textbf {\bibinfo {volume} {77}},\ \bibinfo {pages} {3865} (\bibinfo {year} {1996})}\BibitemShut {NoStop}%
\bibitem [{\citenamefont {Kyrtsos}\ \emph {et~al.}(2017)\citenamefont {Kyrtsos}, \citenamefont {Matsubara},\ and\ \citenamefont {Bellotti}}]{13_Kyrtsos_2017}%
  \BibitemOpen
  \bibfield  {author} {\bibinfo {author} {\bibfnamefont {A.}~\bibnamefont {Kyrtsos}}, \bibinfo {author} {\bibfnamefont {M.}~\bibnamefont {Matsubara}},\ and\ \bibinfo {author} {\bibfnamefont {E.}~\bibnamefont {Bellotti}},\ }\href {https://doi.org/10.1103/PhysRevB.95.245202} {\bibfield  {journal} {\bibinfo  {journal} {Phys. Rev. B}\ }\textbf {\bibinfo {volume} {95}},\ \bibinfo {pages} {245202} (\bibinfo {year} {2017})}\BibitemShut {NoStop}%
\bibitem [{\citenamefont {Yoshioka}\ \emph {et~al.}(2007)\citenamefont {Yoshioka}, \citenamefont {Hayashi}, \citenamefont {Kuwabara}, \citenamefont {Oba}, \citenamefont {Matsunaga},\ and\ \citenamefont {Tanaka}}]{40_Yoshioka_2007}%
  \BibitemOpen
  \bibfield  {author} {\bibinfo {author} {\bibfnamefont {S.}~\bibnamefont {Yoshioka}}, \bibinfo {author} {\bibfnamefont {H.}~\bibnamefont {Hayashi}}, \bibinfo {author} {\bibfnamefont {A.}~\bibnamefont {Kuwabara}}, \bibinfo {author} {\bibfnamefont {F.}~\bibnamefont {Oba}}, \bibinfo {author} {\bibfnamefont {K.}~\bibnamefont {Matsunaga}},\ and\ \bibinfo {author} {\bibfnamefont {I.}~\bibnamefont {Tanaka}},\ }\href {https://doi.org/10.1088/0953-8984/19/34/346211} {\bibfield  {journal} {\bibinfo  {journal} {Journal of Physics: Condensed Matter}\ }\textbf {\bibinfo {volume} {19}},\ \bibinfo {pages} {346211} (\bibinfo {year} {2007})}\BibitemShut {NoStop}%
\bibitem [{\citenamefont {Zacherle}\ \emph {et~al.}(2013)\citenamefont {Zacherle}, \citenamefont {Schmidt},\ and\ \citenamefont {Martin}}]{33_Zacherle_2013}%
  \BibitemOpen
  \bibfield  {author} {\bibinfo {author} {\bibfnamefont {T.}~\bibnamefont {Zacherle}}, \bibinfo {author} {\bibfnamefont {P.~C.}\ \bibnamefont {Schmidt}},\ and\ \bibinfo {author} {\bibfnamefont {M.}~\bibnamefont {Martin}},\ }\href {https://doi.org/10.1103/PhysRevB.87.235206} {\bibfield  {journal} {\bibinfo  {journal} {Phys. Rev. B}\ }\textbf {\bibinfo {volume} {87}},\ \bibinfo {pages} {235206} (\bibinfo {year} {2013})}\BibitemShut {NoStop}%
\bibitem [{\citenamefont {Geller}(1960)}]{34_Geller_1960}%
  \BibitemOpen
  \bibfield  {author} {\bibinfo {author} {\bibfnamefont {S.}~\bibnamefont {Geller}},\ }\href {https://doi.org/10.1063/1.1731237} {\bibfield  {journal} {\bibinfo  {journal} {The Journal of Chemical Physics}\ }\textbf {\bibinfo {volume} {33}},\ \bibinfo {pages} {676} (\bibinfo {year} {1960})}\BibitemShut {NoStop}%
\bibitem [{\citenamefont {{\AA}hman}\ \emph {et~al.}(1996)\citenamefont {{\AA}hman}, \citenamefont {Svensson},\ and\ \citenamefont {Albertsson}}]{35_Ahman_1996}%
  \BibitemOpen
  \bibfield  {author} {\bibinfo {author} {\bibfnamefont {J.}~\bibnamefont {{\AA}hman}}, \bibinfo {author} {\bibfnamefont {G.}~\bibnamefont {Svensson}},\ and\ \bibinfo {author} {\bibfnamefont {J.}~\bibnamefont {Albertsson}},\ }\href {https://doi.org/10.1107/S0108270195016404} {\bibfield  {journal} {\bibinfo  {journal} {Acta Crystallographica Section C}\ }\textbf {\bibinfo {volume} {52}},\ \bibinfo {pages} {1336} (\bibinfo {year} {1996})}\BibitemShut {NoStop}%
\bibitem [{\citenamefont {Monkhorst}\ and\ \citenamefont {Pack}(1976)}]{31_Monkhorst_1976}%
  \BibitemOpen
  \bibfield  {author} {\bibinfo {author} {\bibfnamefont {H.~J.}\ \bibnamefont {Monkhorst}}\ and\ \bibinfo {author} {\bibfnamefont {J.~D.}\ \bibnamefont {Pack}},\ }\href {https://doi.org/10.1103/PhysRevB.13.5188} {\bibfield  {journal} {\bibinfo  {journal} {Phys. Rev. B}\ }\textbf {\bibinfo {volume} {13}},\ \bibinfo {pages} {5188} (\bibinfo {year} {1976})}\BibitemShut {NoStop}%
\bibitem [{\citenamefont {Freysoldt}\ \emph {et~al.}(2014)\citenamefont {Freysoldt}, \citenamefont {Grabowski}, \citenamefont {Hickel}, \citenamefont {Neugebauer}, \citenamefont {Kresse}, \citenamefont {Janotti},\ and\ \citenamefont {Van~de Walle}}]{37_Freysoldt_2014}%
  \BibitemOpen
  \bibfield  {author} {\bibinfo {author} {\bibfnamefont {C.}~\bibnamefont {Freysoldt}}, \bibinfo {author} {\bibfnamefont {B.}~\bibnamefont {Grabowski}}, \bibinfo {author} {\bibfnamefont {T.}~\bibnamefont {Hickel}}, \bibinfo {author} {\bibfnamefont {J.}~\bibnamefont {Neugebauer}}, \bibinfo {author} {\bibfnamefont {G.}~\bibnamefont {Kresse}}, \bibinfo {author} {\bibfnamefont {A.}~\bibnamefont {Janotti}},\ and\ \bibinfo {author} {\bibfnamefont {C.~G.}\ \bibnamefont {Van~de Walle}},\ }\href {https://doi.org/10.1103/RevModPhys.86.253} {\bibfield  {journal} {\bibinfo  {journal} {Rev. Mod. Phys.}\ }\textbf {\bibinfo {volume} {86}},\ \bibinfo {pages} {253} (\bibinfo {year} {2014})}\BibitemShut {NoStop}%
\bibitem [{\citenamefont {Lany}\ and\ \citenamefont {Zunger}(2009)}]{38_Lany_2009}%
  \BibitemOpen
  \bibfield  {author} {\bibinfo {author} {\bibfnamefont {S.}~\bibnamefont {Lany}}\ and\ \bibinfo {author} {\bibfnamefont {A.}~\bibnamefont {Zunger}},\ }\href {https://doi.org/10.1088/0965-0393/17/8/084002} {\bibfield  {journal} {\bibinfo  {journal} {Modelling and Simulation in Materials Science and Engineering}\ }\textbf {\bibinfo {volume} {17}},\ \bibinfo {pages} {084002} (\bibinfo {year} {2009})}\BibitemShut {NoStop}%
\bibitem [{\citenamefont {Adamczyk}\ \emph {et~al.}(2021)\citenamefont {Adamczyk}, \citenamefont {Gomes}, \citenamefont {Qu}, \citenamefont {Rome}, \citenamefont {Baumann}, \citenamefont {Ertekin},\ and\ \citenamefont {Toberer}}]{39_Adamczyk_2021}%
  \BibitemOpen
  \bibfield  {author} {\bibinfo {author} {\bibfnamefont {J.~M.}\ \bibnamefont {Adamczyk}}, \bibinfo {author} {\bibfnamefont {L.~C.}\ \bibnamefont {Gomes}}, \bibinfo {author} {\bibfnamefont {J.}~\bibnamefont {Qu}}, \bibinfo {author} {\bibfnamefont {G.~A.}\ \bibnamefont {Rome}}, \bibinfo {author} {\bibfnamefont {S.~M.}\ \bibnamefont {Baumann}}, \bibinfo {author} {\bibfnamefont {E.}~\bibnamefont {Ertekin}},\ and\ \bibinfo {author} {\bibfnamefont {E.~S.}\ \bibnamefont {Toberer}},\ }\href {https://doi.org/10.1021/acs.chemmater.0c04041} {\bibfield  {journal} {\bibinfo  {journal} {Chemistry of Materials}\ }\textbf {\bibinfo {volume} {33}},\ \bibinfo {pages} {359} (\bibinfo {year} {2021})}\BibitemShut {NoStop}%
\bibitem [{\citenamefont {Henkelman}\ \emph {et~al.}(2000)\citenamefont {Henkelman}, \citenamefont {Uberuaga},\ and\ \citenamefont {J{\'{o}}nsson}}]{Henkelman2000}%
  \BibitemOpen
  \bibfield  {author} {\bibinfo {author} {\bibfnamefont {G.}~\bibnamefont {Henkelman}}, \bibinfo {author} {\bibfnamefont {B.~P.}\ \bibnamefont {Uberuaga}},\ and\ \bibinfo {author} {\bibfnamefont {H.}~\bibnamefont {J{\'{o}}nsson}},\ }\href {https://doi.org/10.1063/1.1329672} {\bibfield  {journal} {\bibinfo  {journal} {Journal of Chemical Physics}\ }\textbf {\bibinfo {volume} {113}},\ \bibinfo {pages} {9901} (\bibinfo {year} {2000})}\BibitemShut {NoStop}%
\bibitem [{\citenamefont {Frodason}\ \emph {et~al.}(2023)\citenamefont {Frodason}, \citenamefont {Krzyzaniak}, \citenamefont {Vines}, \citenamefont {Varley}, \citenamefont {Van~de Walle},\ and\ \citenamefont {Johansen}}]{10_Frodason_2023}%
  \BibitemOpen
  \bibfield  {author} {\bibinfo {author} {\bibfnamefont {Y.~K.}\ \bibnamefont {Frodason}}, \bibinfo {author} {\bibfnamefont {P.~P.}\ \bibnamefont {Krzyzaniak}}, \bibinfo {author} {\bibfnamefont {L.}~\bibnamefont {Vines}}, \bibinfo {author} {\bibfnamefont {J.~B.}\ \bibnamefont {Varley}}, \bibinfo {author} {\bibfnamefont {C.~G.}\ \bibnamefont {Van~de Walle}},\ and\ \bibinfo {author} {\bibfnamefont {K.~M.~H.}\ \bibnamefont {Johansen}},\ }\bibfield  {journal} {\bibinfo  {journal} {APL Materials}\ }\textbf {\bibinfo {volume} {11}},\ \href {https://doi.org/10.1063/5.0142671} {10.1063/5.0142671} (\bibinfo {year} {2023}),\ \bibinfo {note} {041121}\BibitemShut {NoStop}%
\bibitem [{\citenamefont {Blanco}\ \emph {et~al.}(2005)\citenamefont {Blanco}, \citenamefont {Sahariah}, \citenamefont {Jiang}, \citenamefont {Costales},\ and\ \citenamefont {Pandey}}]{14_Blanco_2005}%
  \BibitemOpen
  \bibfield  {author} {\bibinfo {author} {\bibfnamefont {M.~A.}\ \bibnamefont {Blanco}}, \bibinfo {author} {\bibfnamefont {M.~B.}\ \bibnamefont {Sahariah}}, \bibinfo {author} {\bibfnamefont {H.}~\bibnamefont {Jiang}}, \bibinfo {author} {\bibfnamefont {A.}~\bibnamefont {Costales}},\ and\ \bibinfo {author} {\bibfnamefont {R.}~\bibnamefont {Pandey}},\ }\href {https://doi.org/10.1103/PhysRevB.72.184103} {\bibfield  {journal} {\bibinfo  {journal} {Phys. Rev. B}\ }\textbf {\bibinfo {volume} {72}},\ \bibinfo {pages} {184103} (\bibinfo {year} {2005})}\BibitemShut {NoStop}%
\bibitem [{\citenamefont {Lehtom{\"a}ki}\ \emph {et~al.}(2020)\citenamefont {Lehtom{\"a}ki}, \citenamefont {Li},\ and\ \citenamefont {Rinke}}]{Lehtom2020}%
  \BibitemOpen
  \bibfield  {author} {\bibinfo {author} {\bibfnamefont {J.}~\bibnamefont {Lehtom{\"a}ki}}, \bibinfo {author} {\bibfnamefont {J.}~\bibnamefont {Li}},\ and\ \bibinfo {author} {\bibfnamefont {P.}~\bibnamefont {Rinke}},\ }\href {https://doi.org/10.1088/2399-6528/abcd74} {\bibfield  {journal} {\bibinfo  {journal} {Journal of Physics Communications}\ }\textbf {\bibinfo {volume} {4}},\ \bibinfo {pages} {125001} (\bibinfo {year} {2020})}\BibitemShut {NoStop}%
\bibitem [{\citenamefont {Tuttle}\ \emph {et~al.}(2023)\citenamefont {Tuttle}, \citenamefont {Karom}, \citenamefont {O'Hara}, \citenamefont {Schrimpf},\ and\ \citenamefont {Pantelides}}]{Tuttle2023}%
  \BibitemOpen
  \bibfield  {author} {\bibinfo {author} {\bibfnamefont {B.~R.}\ \bibnamefont {Tuttle}}, \bibinfo {author} {\bibfnamefont {N.~J.}\ \bibnamefont {Karom}}, \bibinfo {author} {\bibfnamefont {A.}~\bibnamefont {O'Hara}}, \bibinfo {author} {\bibfnamefont {R.~D.}\ \bibnamefont {Schrimpf}},\ and\ \bibinfo {author} {\bibfnamefont {S.~T.}\ \bibnamefont {Pantelides}},\ }\href {https://doi.org/10.1063/5.0124285} {\bibfield  {journal} {\bibinfo  {journal} {Journal of Applied Physics}\ }\textbf {\bibinfo {volume} {133}},\ \bibinfo {pages} {015703} (\bibinfo {year} {2023})}\BibitemShut {NoStop}%
\bibitem [{\citenamefont {De\'ak}\ \emph {et~al.}(2017)\citenamefont {De\'ak}, \citenamefont {Duy~Ho}, \citenamefont {Seemann}, \citenamefont {Aradi}, \citenamefont {Lorke},\ and\ \citenamefont {Frauenheim}}]{Deak2017}%
  \BibitemOpen
  \bibfield  {author} {\bibinfo {author} {\bibfnamefont {P.}~\bibnamefont {De\'ak}}, \bibinfo {author} {\bibfnamefont {Q.}~\bibnamefont {Duy~Ho}}, \bibinfo {author} {\bibfnamefont {F.}~\bibnamefont {Seemann}}, \bibinfo {author} {\bibfnamefont {B.}~\bibnamefont {Aradi}}, \bibinfo {author} {\bibfnamefont {M.}~\bibnamefont {Lorke}},\ and\ \bibinfo {author} {\bibfnamefont {T.}~\bibnamefont {Frauenheim}},\ }\href {https://doi.org/10.1103/PhysRevB.95.075208} {\bibfield  {journal} {\bibinfo  {journal} {Phys. Rev. B}\ }\textbf {\bibinfo {volume} {95}},\ \bibinfo {pages} {075208} (\bibinfo {year} {2017})}\BibitemShut {NoStop}%
\bibitem [{\citenamefont {Sun}\ \emph {et~al.}(2019)\citenamefont {Sun}, \citenamefont {Gao}, \citenamefont {Xue},\ and\ \citenamefont {Zhao}}]{Sun2019}%
  \BibitemOpen
  \bibfield  {author} {\bibinfo {author} {\bibfnamefont {D.}~\bibnamefont {Sun}}, \bibinfo {author} {\bibfnamefont {Y.}~\bibnamefont {Gao}}, \bibinfo {author} {\bibfnamefont {J.}~\bibnamefont {Xue}},\ and\ \bibinfo {author} {\bibfnamefont {J.}~\bibnamefont {Zhao}},\ }\href {https://doi.org/https://doi.org/10.1016/j.jallcom.2019.04.253} {\bibfield  {journal} {\bibinfo  {journal} {Journal of Alloys and Compounds}\ }\textbf {\bibinfo {volume} {794}},\ \bibinfo {pages} {374} (\bibinfo {year} {2019})}\BibitemShut {NoStop}%
\bibitem [{\citenamefont {Li}\ \emph {et~al.}(2024)\citenamefont {Li}, \citenamefont {Zhao}, \citenamefont {Li}, \citenamefont {Guan}, \citenamefont {Jia}, \citenamefont {Lin},\ and\ \citenamefont {Zhao}}]{Li2024}%
  \BibitemOpen
  \bibfield  {author} {\bibinfo {author} {\bibfnamefont {Q.}~\bibnamefont {Li}}, \bibinfo {author} {\bibfnamefont {J.}~\bibnamefont {Zhao}}, \bibinfo {author} {\bibfnamefont {Y.}~\bibnamefont {Li}}, \bibinfo {author} {\bibfnamefont {X.}~\bibnamefont {Guan}}, \bibinfo {author} {\bibfnamefont {Z.}~\bibnamefont {Jia}}, \bibinfo {author} {\bibfnamefont {N.}~\bibnamefont {Lin}},\ and\ \bibinfo {author} {\bibfnamefont {X.}~\bibnamefont {Zhao}},\ }\href {https://doi.org/10.1021/acs.jpcc.4c01787} {\bibfield  {journal} {\bibinfo  {journal} {The Journal of Physical Chemistry C}\ }\textbf {\bibinfo {volume} {128}},\ \bibinfo {pages} {11817} (\bibinfo {year} {2024})}\BibitemShut {NoStop}%
\bibitem [{\citenamefont {Martins}\ \emph {et~al.}(2024)\citenamefont {Martins}, \citenamefont {de~Campos~Viana}, \citenamefont {das Gra{\c c}as~Santos}, \citenamefont {Borges}, \citenamefont {Welch}, \citenamefont {Borges},\ and\ \citenamefont {Scolfaro}}]{Martins2024}%
  \BibitemOpen
  \bibfield  {author} {\bibinfo {author} {\bibfnamefont {N.~R.}\ \bibnamefont {Martins}}, \bibinfo {author} {\bibfnamefont {L.~A.~F.}\ \bibnamefont {de~Campos~Viana}}, \bibinfo {author} {\bibfnamefont {A.~A.}\ \bibnamefont {das Gra{\c c}as~Santos}}, \bibinfo {author} {\bibfnamefont {D.~D.}\ \bibnamefont {Borges}}, \bibinfo {author} {\bibfnamefont {E.}~\bibnamefont {Welch}}, \bibinfo {author} {\bibfnamefont {P.~D.}\ \bibnamefont {Borges}},\ and\ \bibinfo {author} {\bibfnamefont {L.}~\bibnamefont {Scolfaro}},\ }\href {https://doi.org/10.1116/6.0003435} {\bibfield  {journal} {\bibinfo  {journal} {Journal of Vacuum Science \& Technology A}\ }\textbf {\bibinfo {volume} {42}},\ \bibinfo {pages} {032801} (\bibinfo {year} {2024})}\BibitemShut {NoStop}%
\end{thebibliography}%

\end{document}


\newpage
\renewcommand\thetable{S\arabic{table}} 
\newcounter{table1} 
\begin{table}[!hbtp]
\centering

\small
\renewcommand{\arraystretch}{1.2}
    \centering
    \begin{tabular}{c c c c c c}
    \hline
    \hline
    \multicolumn{1}{c}{Principle hop (PH)} & \multicolumn{1}{c}{Transformation}  & \multicolumn{1}{c}{Distance (\AA)} & \multicolumn{1}{c}{$\rightharpoonup$ (eV)} & \multicolumn{1}{c}{$\leftharpoondown$ (eV)} \\ \hline
        1 & $\text{O}_\text{ia}^0\rightleftharpoons\text{O}_\text{ia}^0$ & 2.95 & 0.85 & 0.85 \\ 
        2 & $\text{O}_\text{ia}^0\rightleftharpoons\text{O}_\text{ia}^0$ & 0.14 & 0.08 & 0.08 \\ 
        3 & $\text{O}_\text{ia}^0\rightleftharpoons\text{O}_\text{ib}^0$ & 2.89 & 1.10 & 0.35 \\ 
        4 & $\text{O}_\text{ia}^0\rightleftharpoons\text{O}_\text{id}^0$ & 2.80 & 0.88 & 0.65 \\ 
        5 & $\text{O}_\text{ia}^0\rightleftharpoons\text{O}_\text{if}^0$ & 0.42 & 0.27 & 0.22 \\ 
        6 & $\text{O}_\text{ia}^0\rightleftharpoons\text{O}_\text{if}^0$ & 2.92 & 0.3 & 0.25 \\ 
        7 & $\text{O}_\text{ib}^0\rightleftharpoons\text{O}_\text{ib}^0$ & 2.62 & 0.50 & 0.50 \\ 
        8 & $\text{O}_\text{ib}^0\rightleftharpoons\text{O}_\text{ic}^0$ & 0.26 & 0.39 & 0.85 \\ 
        9 & $\text{O}_\text{ib}^0\rightleftharpoons\text{O}_\text{ie}^0$ & 2.63 & 1.17 & 1.71 \\ 
        10 & $\text{O}_\text{ic}^0\rightleftharpoons\text{O}_\text{ic}^0$ & 2.58 & 0.59 & 0.59 \\ 
        11 & $\text{O}_\text{ic}^0\rightleftharpoons\text{O}_\text{ie}^0$ & 2.57 & 0.39 & 0.88 \\ 
        12 & $\text{O}_\text{id}^0\rightleftharpoons\text{O}_\text{ie}^0$ & 0.35 & 0.31 & 0.40 \\ 
        13 & $\text{O}_\text{id}^0\rightleftharpoons\text{O}_\text{ie}^0$ & 3.11 & 1.22 & 1.31 \\ 
        14 & $\text{O}_\text{id}^0\rightleftharpoons\text{O}_\text{if}^0$ & 2.80 & 1.41 & 1.59 \\ 
        15 & $\text{O}_\text{ie}^0\rightleftharpoons\text{O}_\text{ie}^0$ & 3.31 & 1.34 & 1.34 \\ 
        16 & $\text{O}_\text{ie}^0\rightleftharpoons\text{O}_\text{ie}^0$ & 3.08 & 0.88 & 0.88 \\ 
        17 & $\text{O}_\text{if}^0\rightleftharpoons\text{O}_\text{if}^0$ & 2.71 & 0.25 & 0.25 \\ 
        \hline
        \hline
        \end{tabular}
\caption{A list of principle hops (PHs), which cannot be further decomposed into multiple shorter elementary hops, between $\text{O}_\text{i}^{0}$ configurations and their corresponding migration barriers in forward ($\rightharpoonup$) and backward ($\leftharpoondown$) directions. The migration barriers were calculated by the ci-NEB method using the PBE functional as described in Section II.}
\end{table}

\newpage
\begin{table}[!hbtp]
\centering
\small
\renewcommand{\arraystretch}{1.2}
    \centering
    \begin{tabular}{c c c c c c}
    \hline
    \hline
        \multicolumn{1}{c}{Principle hop (PH)} & \multicolumn{1}{c}{Transformation}  & \multicolumn{1}{c}{Distance (\AA)} & \multicolumn{1}{c}{$\rightharpoonup$ (eV)} & \multicolumn{1}{c}{$\leftharpoondown$ (eV)} \\ \hline
        18 & $\text{O}_\text{ig}^{2-}\rightleftharpoons\text{O}_\text{ig}^{2-}$ & 2.14 & 0.24 & 0.24 \\ 
        19 & $\text{O}_\text{ig}^{2-}\rightleftharpoons\text{O}_\text{ig}^{2-}$ & 3.09 & 2.18 & 2.18 \\ 
        20 & $\text{O}_\text{ig}^{2-}\rightleftharpoons\text{O}_\text{ih}^{2-}$ & 1.45 & 1.12 & 0.61 \\
        21 & $\text{O}_\text{ig}^{2-}\rightleftharpoons\text{O}_\text{ih}^{2-}$ & 1.64 & 0.71 & 0.19 \\
        22 & $\text{O}_\text{ig}^{2-}\rightleftharpoons\text{O}_\text{ih}^{2-}$ & 3.32 & 0.71 & 0.20 \\ 
        23 & $\text{O}_\text{ig}^{2-}\rightleftharpoons\text{O}_\text{ih}^{2-}$ & 3.50 & 2.76 & 2.24 \\ 
        24 & $\text{O}_\text{ig}^{2-}\rightleftharpoons\text{O}_\text{ii}^{2-}$ & 3.88 & 2.72 & 1.50 \\ 
        25 & $\text{O}_\text{ig}^{2-}\rightleftharpoons\text{O}_\text{ij}^{2-}$ & 3.49 & 3.36 & 2.29 \\ 
        26 & $\text{O}_\text{ig}^{2-}\rightleftharpoons\text{O}_\text{ik}^{2-}$ & 3.92 & 1.44 & 0.76 \\ 
        27 & $\text{O}_\text{ih}^{2-}\rightleftharpoons\text{O}_\text{ih}^{2-}$ & 3.09 & 0.33 & 0.33 \\ 
        28 & $\text{O}_\text{ih}^{2-}\rightleftharpoons\text{O}_\text{ih}^{2-}$ & 3.53 & 2.99 & 2.99 \\ 
        29 & $\text{O}_\text{ih}^{2-}\rightleftharpoons\text{O}_\text{ih}^{2-}$ & 3.72 & 2.96 & 2.96 \\ 
        30 & $\text{O}_\text{ih}^{2-}\rightleftharpoons\text{O}_\text{ii}^{2-}$ & 1.88 & 0.80 & 0.10 \\ 
        31 & $\text{O}_\text{ih}^{2-}\rightleftharpoons\text{O}_\text{ij}^{2-}$ & 3.80 & 2.63 & 2.09 \\ 
        32 & $\text{O}_\text{ih}^{2-}\rightleftharpoons\text{O}_\text{ik}^{2-}$ & 1.59 & 0.47 & 0.30 \\ 
        33 & $\text{O}_\text{ih}^{2-}\rightleftharpoons\text{O}_\text{ik}^{2-}$ & 2.27 & 0.71 & 0.54 \\ 
        34 & $\text{O}_\text{ii}^{2-}\rightleftharpoons\text{O}_\text{ii}^{2-}$ & 1.31 & 0.72 & 0.72 \\ 
        35 & $\text{O}_\text{ii}^{2-}\rightleftharpoons\text{O}_\text{ii}^{2-}$ & 1.78 & 0.74 & 0.74 \\ 
        36 & $\text{O}_\text{ii}^{2-}\rightleftharpoons\text{O}_\text{ik}^{2-}$ & 0.66 & 0.00 & 0.54 \\ 
        37 & $\text{O}_\text{ii}^{2-}\rightleftharpoons\text{O}_\text{ik}^{2-}$ & 1.37 & 0.36 & 0.90 \\ 
        38 & $\text{O}_\text{ij}^{2-}\rightleftharpoons\text{O}_\text{ij}^{2-}$ & 1.52 & 0.79 & 0.79 \\ 
        39 & $\text{O}_\text{ij}^{2-}\rightleftharpoons\text{O}_\text{ij}^{2-}$ & 1.57 & 0.66 & 0.66 \\ 
        40 & $\text{O}_\text{ij}^{2-}\rightleftharpoons\text{O}_\text{ik}^{2-}$ & 2.85 & 0.15 & 0.53 \\ 
        41 & $\text{O}_\text{ik}^{2-}\rightleftharpoons\text{O}_\text{ik}^{2-}$ & 1.10 & 0.30 & 0.30 \\ 
        42 & $\text{O}_\text{ik}^{2-}\rightleftharpoons\text{O}_\text{ik}^{2-}$ & 1.30 & 0.59 & 0.59 \\ 
        43 & $\text{O}_\text{ik}^{2-}\rightleftharpoons\text{O}_\text{ik}^{2-}$ & 1.70 & 1.28 & 1.28 \\ 
        44 & $\text{O}_\text{ik}^{2-}\rightleftharpoons\text{O}_\text{ik}^{2-}$ & 1.99 & 0.84 & 0.84 \\ 
        45 & $\text{O}_\text{ik}^{2-}\rightleftharpoons\text{O}_\text{ik}^{2-}$ & 3.09 & 0.86 & 0.86 \\ 
        \hline
        \hline
        \end{tabular}
\caption{A list of principle hops (PHs), which cannot be further decomposed into multiple shorter elementary hops, between $\text{O}_\text{i}^{2-}$ configurations and their corresponding migration barriers in forward ($\rightharpoonup$) and backward ($\leftharpoondown$) directions. The migration barriers were calculated by the ci-NEB method using the PBE functional as described in Section II.}
\end{table}

\newpage
\begin{figure*}[!hbtp]
\centering
\includegraphics[width=1\linewidth]{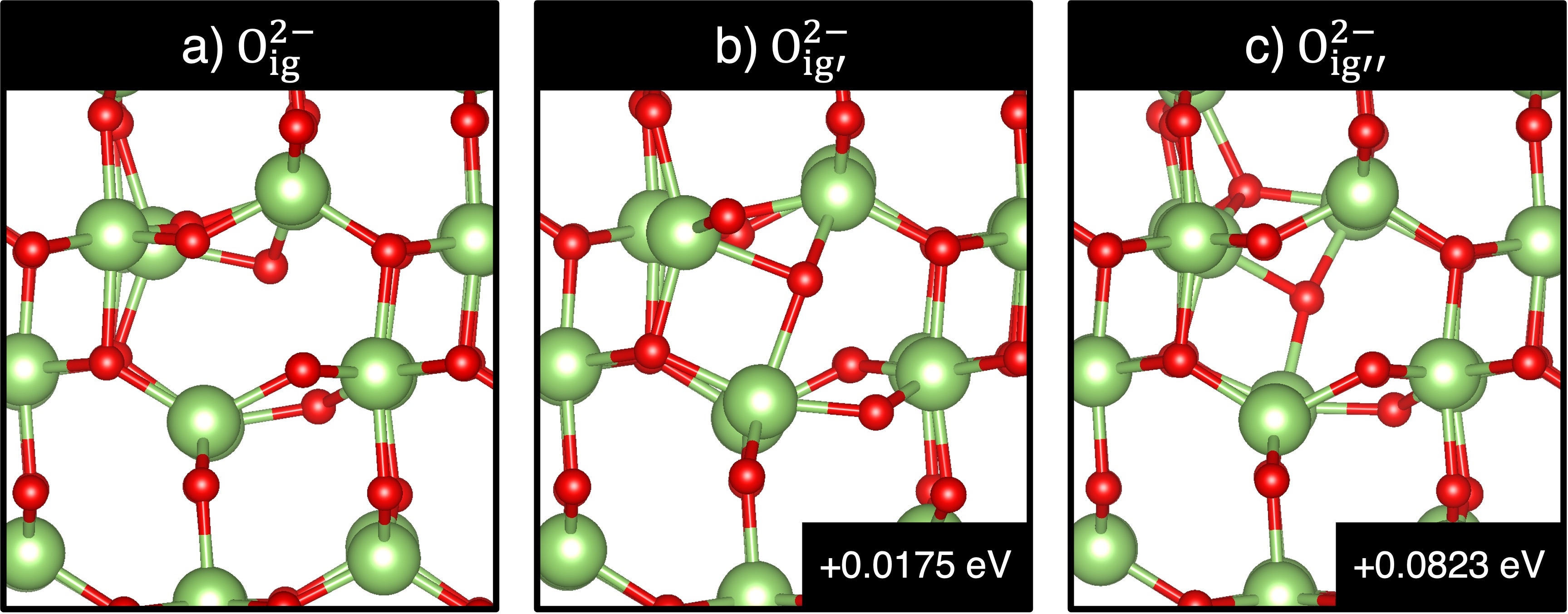}
\caption{Configurations of (a) $\text{O}_{\text{ig}}^{2-}$, (b) $\text{O}_{\text{ig'}}^{2-}$, and (c) $\text{O}_{\text{ig''}}^{2-}$ within the A channel. Energies given in the bottom right corners of (b) and (c) are relative to E$_\text{for}[\text{O}_{\text{ig}}^{2-}]$.}
\end{figure*}

\newpage
\begin{figure*}[!hbtp]
     \centering
    \includegraphics[width=1\linewidth]{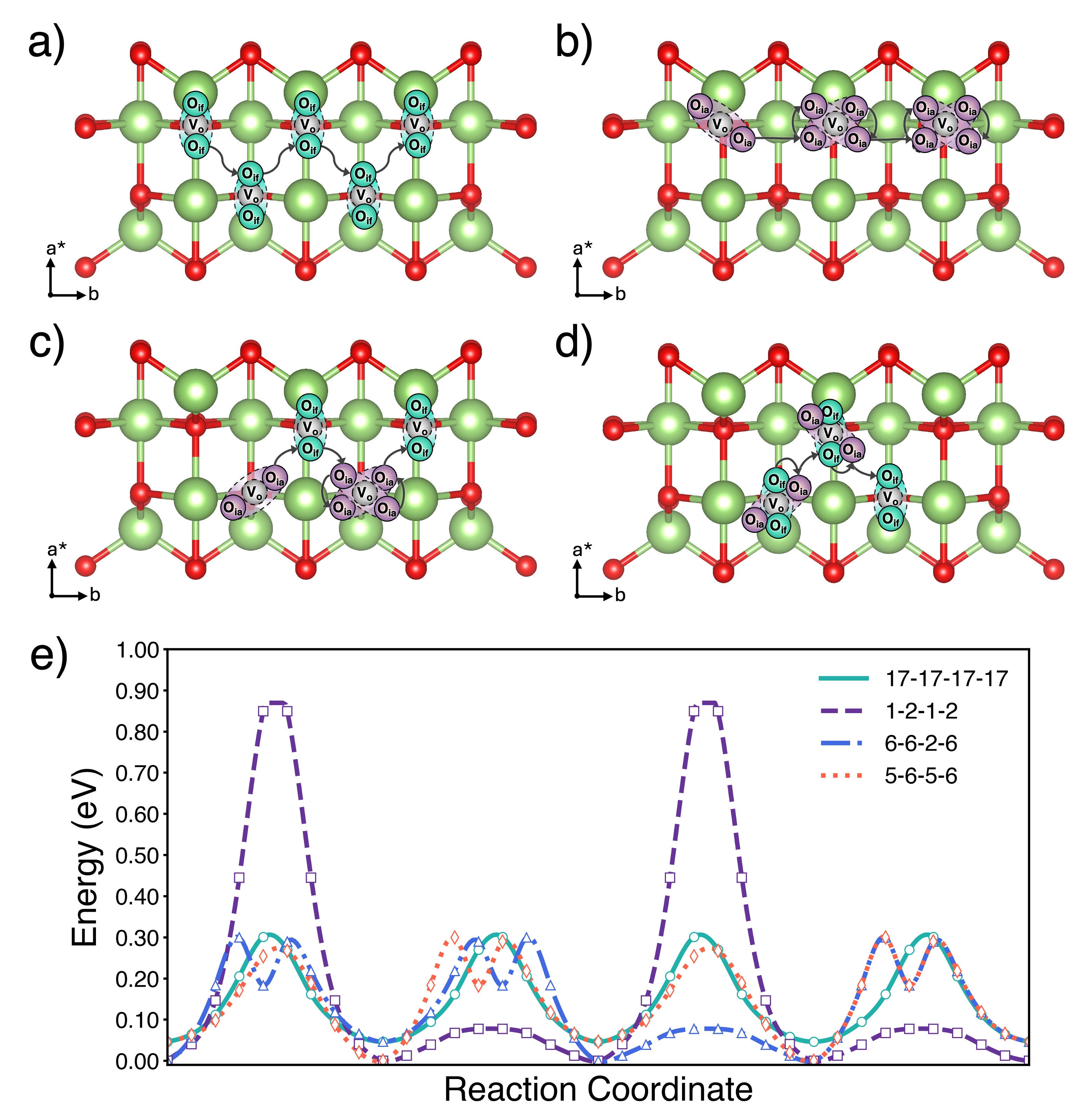}
    \caption{(a-d) Schematic representation of some possible (010) diffusion pathways of neutral ($q=0$) oxygen interstitials and (e) corresponding energies along the energy minimum pathways. The pathways consist of principle hops (PHs) (a) 17-17-17-17, (b) 1-2-1-2 (c) 6-6-2-6, and (d) 5-6-5-6.}
\end{figure*}

\newpage
\begin{figure*}[!hbtp]
    \centering
    \includegraphics[width=1\linewidth]{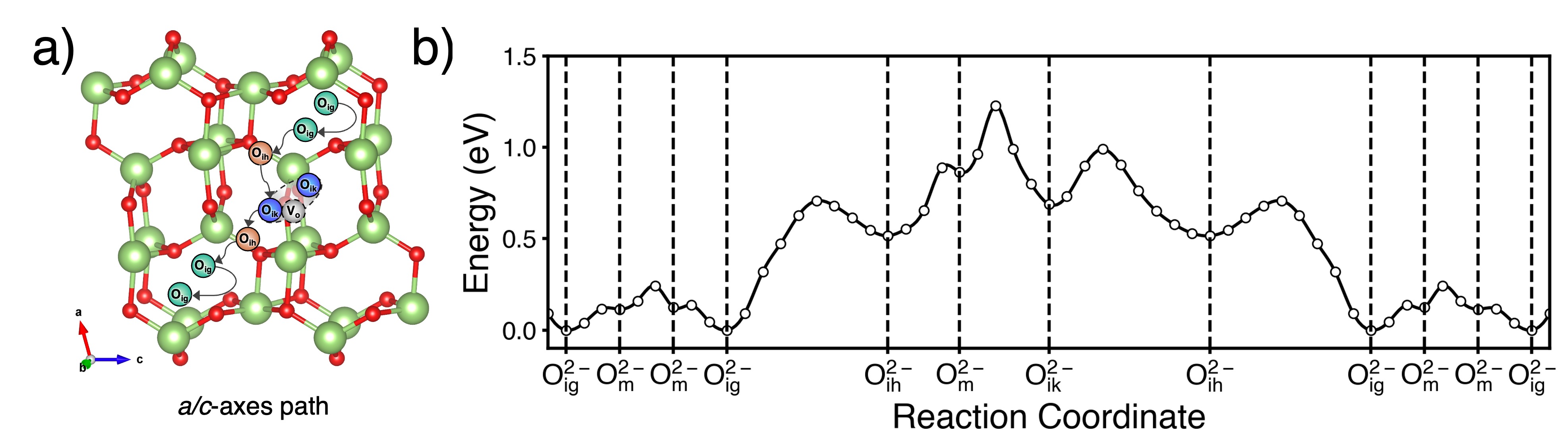}
    \caption{(a) Schematic representation of the dominant [100] and [001] diffusion pathway of charged ($q=2-$) oxygen interstitials and (b) corresponding energies along the energy minimum pathway. Dashed lines in (b) indicate a configuration in the network or intermediate configuration ($\text{O}_{\text{ig'}}$, $\text{O}_{\text{ik'}}$).}
    \label{fig:enter-label}
\end{figure*}

\begin{figure*}[!hbtp]
    \centering
    \includegraphics[width=1\linewidth]{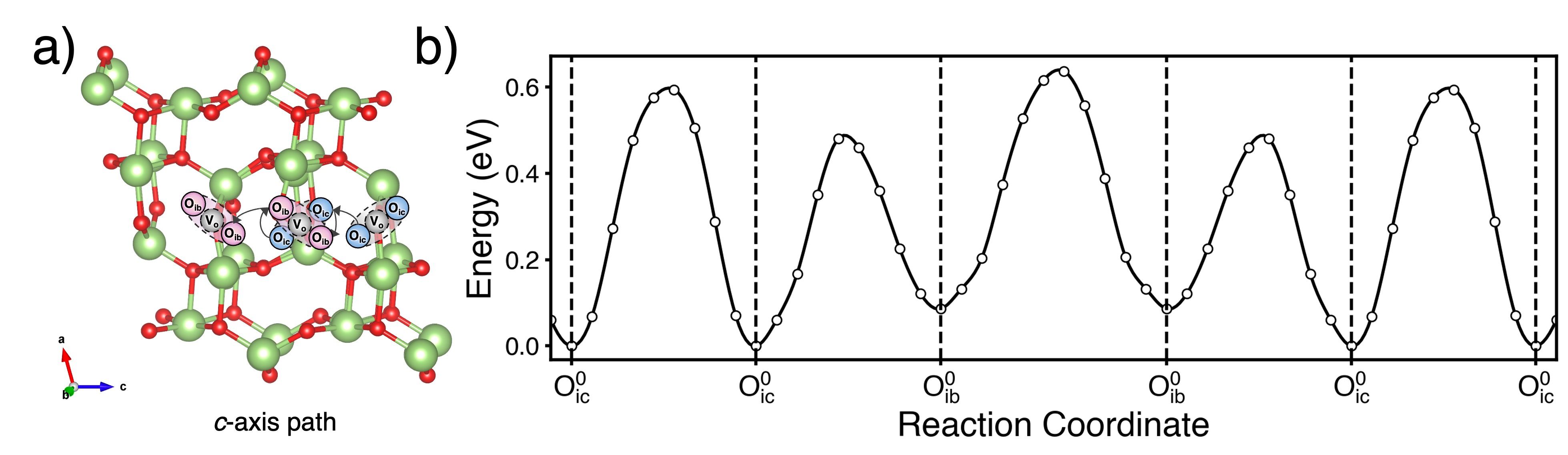}
    \caption{(a) Schematic representation of the dominant [001] diffusion pathway of neutral ($q=0$) oxygen interstitials and (b) corresponding energies along the energy minimum pathway. Dashed lines in (b) indicate configurations in the network.}
\end{figure*}

\begin{figure*}[!hbtp]
    \centering
    \includegraphics[width=1\linewidth]{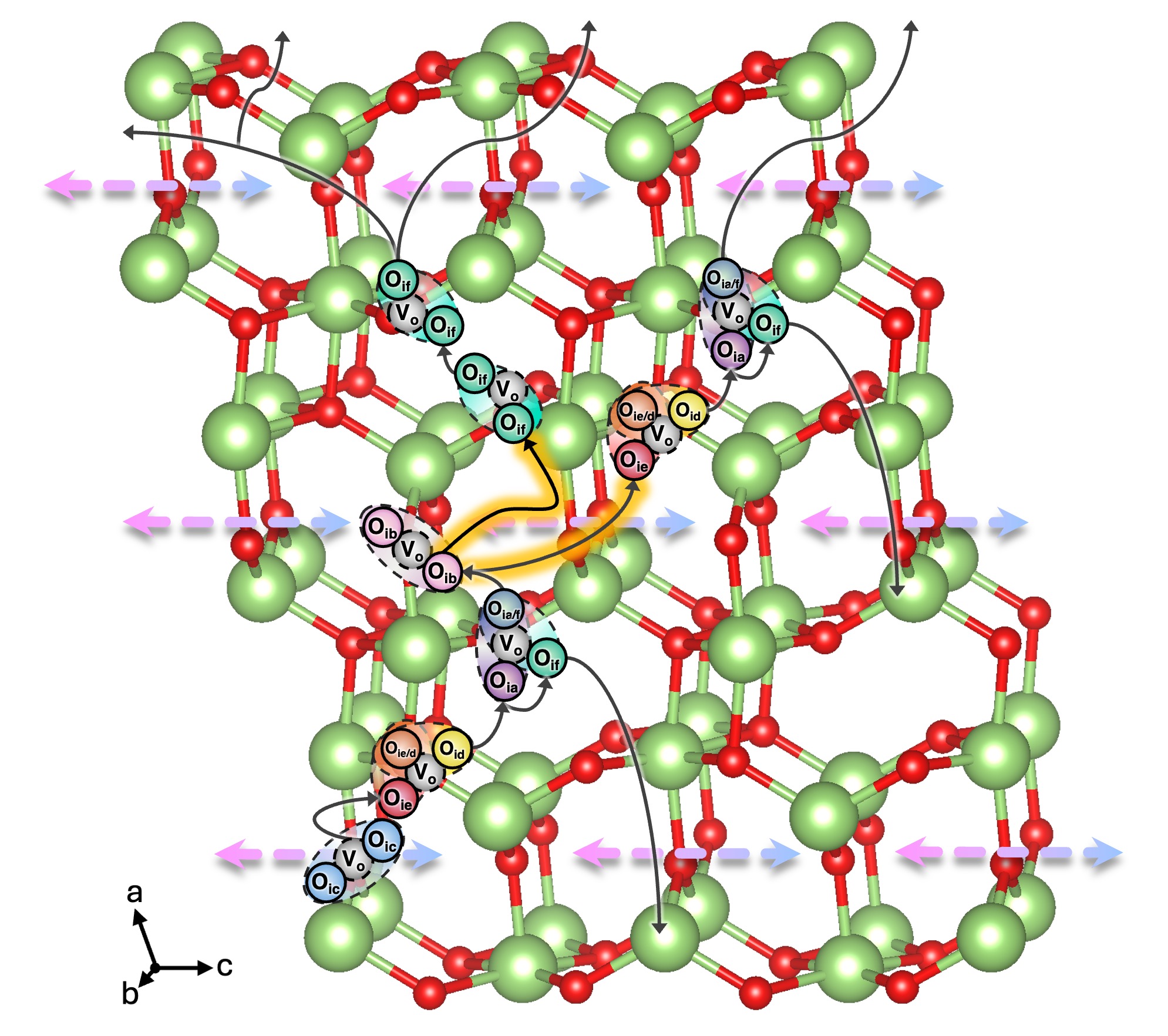}
    \caption{Schematic representation of the dominant [100] diffusion pathway of neutral ($q=0$) oxygen interstitials. Black arrows show potential pathways of the neutral interstitial. The dashed double arrows represent the availability of a [001] hop along the diffusion pathway in Figure S4. Highlighted black arrows emphasize necessary [001] hops along the [001] pathways.}
\end{figure*}
\newpage

\section{Oxygen Chemical Potential Boundaries}
To establish appropriate chemical potential boundaries for the oxygen chemical potential $\mu_{\text{O}}$, we use
\small{
\begin{equation}
  \mu_{\text{O}} = \mu_{\text{O}}^0(T) + \Delta \mu_{\text{O}}(T),  
\end{equation}
}
where $\Delta \mu_{\text{O}}(T)$ represents the deviation from oxygen rich conditions ($\Delta \mu_{\text{O}}(T) = 0$). 
At the other extreme, oxygen poor conditions, $\Delta \mu_{\text{O}}(T) = (1/3) \Delta H_f (\beta\text{-Ga}_2\text{O}_3)$ 
where $\Delta H_f (\beta\text{-Ga}_2\text{O}_3) = -10.78$ eV/fu is the computed E$_\text{f}$ of $\beta$-Ga$_2$O$_3$ (the negative value indicates favorable Ga$_2$O$_3$ formation). 
The reference state $\mu_{\text{O}}^0(T)$ is defined as half of the energy of an $\text{O}_2$ molecule.
This reference is obtained from DFT calculations ($T = 0$ K), and with a correction term to account for the PBE overbinding of the molecule \cite{Patton1997}. 
The region enclosed in dotted lines varies from Ga-rich (low $\text{p(O}_2)$) to O-rich (high $\text{p(O}_2)$), with the dashed midpoint for reference. 
These chemical potential boundaries ensure that the calculated defect E$_\text{f}$ are consistent with the thermodynamic limits of the system.